\documentclass[fleqn,usenatbib]{mnras}

\usepackage[T1]{fontenc}
\usepackage{url}
\DeclareRobustCommand{\VAN}[3]{#2}
\let\VANthebibliography\thebibliography
\def\thebibliography{\DeclareRobustCommand{\VAN}[3]{##3}\VANthebibliography}

\usepackage{graphicx}	%
\usepackage{amsmath}	%
\usepackage{amssymb}	%
\usepackage[dvipsnames]{xcolor} 
\usepackage{newtxtext,newtxmath}
\usepackage[normalem]{ulem}
\usepackage{widetext}

\newcommand{\LIGOlabMIT}{LIGO Laboratory, Massachusetts Institute of Technology, 185 Albany St, Cambridge, MA 02139, USA}
\newcommand{\MKI}{Department of Physics and Kavli Institute for Astrophysics and Space Research, Massachusetts Institute of Technology, \\ 77 Massachusetts Ave, Cambridge, MA 02139, USA}
\newcommand{\CITA}{Canadian Institute for Theoretical Astrophysics, University of Toronto, Toronto, Ontario M5S 3H8, Canada}

\title[NSBH populations and EM prospects]{Population properties and multimessenger prospects of neutron star-black hole mergers following GWTC-3}

\author[Biscoveanu et al.]{
Sylvia Biscoveanu,$^{1,2}$\thanks{sbisco@mit.edu}
Philippe Landry,$^{3}$
Salvatore Vitale$^{1,2}$
\\
$^{1}$\LIGOlabMIT\\
$^{2}$\MKI\\
$^{3}$\CITA
}

\date{Accepted 2022 October 17. Received 2022 October 16; in original form 2022 August 9}

\pubyear{2022}

\begin{document}
\label{firstpage}
\pagerange{\pageref{firstpage}--\pageref{lastpage}}
\maketitle

\begin{abstract}
Neutron star-black hole (NSBH) mergers detected in gravitational waves have the potential to shed light on supernova physics, the dense matter equation of state, and the astrophysical processes that power their potential electromagnetic counterparts. We use the population of four candidate NSBH events detected in gravitational waves so far with a false alarm rate $\leq 1~\mathrm{yr}^{-1}$ to constrain the mass and spin distributions and multimessenger prospects of these systems. We find that the black holes in NSBHs are both less massive and have smaller dimensionless spins than those in black hole binaries. We also find evidence for a mass gap between the most massive neutron stars and least massive black holes in NSBHs at 98.6\% credibility. %
Using an approach driven by gravitational-wave data rather than binary simulations, we find that fewer than 14\% of NSBH mergers detectable in gravitational waves will have an electromagnetic counterpart. While the inferred presence of a mass gap and fraction of sources with a counterpart depend on the event selection and prior knowledge of source classification, the conclusion that the black holes in NSBHs have lower masses and smaller spin parameters than those in black hole binaries is robust.
Finally, we propose a method for the multimessenger analysis of NSBH mergers based on the nondetection of an electromagnetic counterpart and conclude that, even in the most optimistic case, the constraints on the neutron star equation of state that can be obtained with multimessenger NSBH detections are not competitive with those from gravitational-wave measurements of tides in binary neutron star mergers and radio and X-ray pulsar observations.
 
\end{abstract}


\section{Introduction}
The detection of the neutron star-black hole (NSBH) mergers GW200105 and GW200115~\citep{LIGOScientific:2021qlt} by the LIGO~\citep{LIGOScientific:2014pky} and Virgo~\citep{VIRGO:2014yos} gravitational-wave (GW) observatories confirmed the existence of this class of sources and has heralded the study of their population properties~\citep{LIGOScientific:2021psn, Tang:2021bnp, Farah:2021qom, Landry:2021hvl, Zhu:2021jbw, Ye:2022qoe}. Their inferred intrinsic properties and merger rates are consistent with the broad range of predictions for the astrophysical population of neutron-star black hole mergers~\citep{Chattopadhyay:2020lff,Broekgaarden:2021iew, Mandel:2021smh}, although isolated binary evolution likely dominates among the potential formation channels for NSBH systems based on rate arguments~\citep{Mandel:2021smh}. The chirp masses of these systems are expected to lie between $\sim 1.5-5~M_{\odot}$ with a peak around $3~M_{\odot}$, but the shapes and widths of the individual neutron star (NS) and black hole (BH) mass distributions for NSBH are much more uncertain theoretically~\citep{Chattopadhyay:2020lff,Broekgaarden:2021iew}. If the black hole forms second among the two compact objects, it can acquire spin through tidal spin-up of its progenitor, depending on the orbital separation prior to the second supernova~\citep{Qin:2018vaa, Bavera:2021evk, Chattopadhyay:2020lff, Hu:2022ubh}. This is expected for up to $\sim20\%$ of the intrinsic population of NSBH mergers~\citep{Broekgaarden:2021iew, Chattopadhyay:2022cnp}, while systems where the black hole forms first are expected to have negligible spin due to efficient angular momentum transport~\citep{Spruit:2001tz, 2019MNRAS.485.3661F, Fuller:2019sxi}.

NSBH mergers are also potential multimessenger sources if the neutron star gets tidally disrupted outside the black hole innermost stable circular orbit~\citep{Pannarale:2010vs, Foucart:2012nc, Foucart:2018rjc,Pankow:2019oxl}. In this case, the disrupted material can power a range of electromagnetic counterparts including a kilonova~\citep{Tanaka:2013ana, Tanaka:2013ixa, Fernandez:2016sbf, Kawaguchi:2016ana} and short gamma-ray burst (GRB) jet~\citep{1993Natur.361..236M, Janka:1999qu, Paschalidis:2014qra, Shapiro:2017cny, Ruiz:2018wah}. The detection of an electromagnetic counterpart to a NSBH merger observed in gravitational waves (or lack thereof) can be used to place multimessenger constraints on the neutron star equation of state (EoS), remnant mass, and the properties of the kilonova and GRB jet~\citep{Ascenzi:2018mwp, Barbieri:2019sjc, Hinderer:2018pei, Coughlin:2018fis, Coughlin:2019kqf,Chase:2021ood, Raaijmakers:2021slr, Sarin:2022cmu}. Direct measurements of the EoS via gravitational-wave constraints on the tidal deformability of the neutron star in the binary are particularly difficult for NSBH systems---especially those with low signal-to-noise ratios (SNRs) and unequal mass ratios~\citep{Lackey:2011vz, Huang:2020pba}---so taking advantage of the multimessenger information encoded in these systems offers an alternative approach to constrain the EoS.

The latest catalog of gravitational-wave sources (GWTC-3) published by the LIGO-Virgo-KAGRA (LVK) collaboration includes 69 binary black hole (BBH) mergers, four NSBHs, and two binary neutron star (BNS) mergers detected with false alarm rate (FAR) less than 1 per year~\citep{LIGOScientific:2021djp}. GW190814 also meets this FAR criterion, although its source characterization is uncertain as the secondary object is either the most massive neutron star or least massive black hole detected to date~\citep{LIGOScientific:2020zkf}. Previous works have sought both to compare the population properties of the black holes and neutron stars in NSBH mergers to those in BBH or BNS mergers and to constrain the properties of the compact object population as a whole. While the BBH primary mass distribution spans the mass range between $\sim 5-80~M_{\odot}$, \citet{Zhu:2021jbw} find that the masses of the black holes in NSBHs only extend out to $\sim 50~M_{\odot}$ when considering a population of five NSBHs, including one source detected with $\mathrm{FAR} > 1\,\mathrm{yr}^{-1}$, GW191219\_163120. Similarly, the distributions of the effective aligned and precessing spins of NSBHs are found to favor smaller values than those of BBHs. Taking GW190814 to be a NSBH merger with a spinning neutron star, \citet{Ye:2022qoe} find evidence for a mass gap between the lightest black holes and heaviest neutron stars, although \citet{LIGOScientific:2021psn, Farah:2021qom} find that this evidence weakens when considering the full population of compact objects regardless of source type. 

Previous works have also predicted the fraction of NSBH mergers that are expected to be accompanied by an electromagnetic counterpart, $f_{\text{EM-bright}}$, although none have done so by simultaneously fitting for and marginalizing over the binary mass and spin distributions and by accounting for the uncertainty in the NS EoS. \citet{Drozda:2020qab}, \citet{Roman-Garza:2020uou}, \citet{Broekgaarden:2021iew}, and \citet{Fragione:2021cvv} all find that NSBH mergers are unlikely sources of electromagnetic radiation based on population synthesis simulations of NSBH formation via isolated binary evolution; however, $f_{\text{EM-bright}}$ varies across their studies from $\sim 10^{-2}-0.7$ depending on the binary evolution parameters. Higher $f_{\text{EM-bright}}$ is expected for higher black hole spin aligned to the orbital angular momentum and stiffer equations of state, both of which are qualitatively disfavored by current gravitational-wave observations. \citet{Chen:2021fro} constrain the contribution of NSBH mergers to r-process nucleosynthesis based on the observed populations of galactic neutron stars and binary black holes, finding that BNSs contribute at least twice as much to r-process element production.

In this work, we measure the population properties of NSBH sources first using gravitational-wave data alone, focusing on the pairing function determining the distribution of the mass ratio between the neutron star and the black hole. We use the posteriors on the population hyperparameters to take a data-driven approach to estimating $f_{\text{EM-bright}}$ by marginalizing over the uncertainty in the mass and spin distributions along with the EoS. We present the methods and results for the gravitational-wave-only analysis in Section~\ref{sec:gw_only}. Since we find that the $f_{\text{EM-bright}}$ posterior peaks strongly at $f_{\text{EM-bright}}=0$, we next extend the analysis to determine what constraints can be placed on the neutron star EoS under the best-case assumption that no electromagnetic counterpart was identified for any of the four NSBHs in our observed gravitational-wave population because none was produced, presenting the method and results in Section~\ref{sec:joint_analysis}. We conclude with a discussion of the caveats and astrophysical implications of our results in Section~\ref{sec:discussion}.

\section{Gravitational-wave-only analysis of NSBH population properties}
\label{sec:gw_only}
\subsection{Methods}
\label{sec:gw_methods}
We first employ the framework of hierarchical Bayesian inference to measure the mass and spin distributions of NSBH mergers using only gravitational-wave data. Our population includes the four NSBH events reported in GWTC-3 detected with $\mathrm{FAR} \leq 1~\mathrm{yr}^{-1}$: GW190426\_152155, GW190917\_114630, GW200105\_162426, and GW200115\_042309, listed in Table~\ref{tab:events}\footnote{This is the same FAR threshold applied by the LVK in the binary black hole analyses presented in \citet{LIGOScientific:2021psn}, although that work used a threshold of $\mathrm{FAR} \leq 0.25~\mathrm{yr}^{-1}$ for analyses with events containing neutron stars. We exclude the candidate NSBH event GW190531\_023648 reported in \cite{LIGOScientific:2021usb} with maximum $\mathrm{FAR}=0.41~\mathrm{yr}^{-1}$, as parameter estimates are not available. This is consistent with the treatment of this event in \cite{LIGOScientific:2021psn}.}. We do not include GW190814 as the secondary, less massive object in the binary is likely too massive to be a neutron star~\citep{LIGOScientific:2020zkf,EssickLandry2020}, and it is a known outlier relative to observed NSBH (and BBH) systems~\citep{LIGOScientific:2021psn,LIGOScientific:2020kqk}. We seek to obtain posteriors for the hyperparameters $\boldsymbol{\Lambda}_{\mathrm{GW}}$ governing the population-level distributions of binary parameters like the masses and spins, $\boldsymbol{\theta}$, given a set of $N$ events with data $\{d\}$,
\begin{align}
    \label{eq:hyper-post}
    p(\boldsymbol{\Lambda}_{\mathrm{GW}} | \{d\}) \propto p(\{d\} | \boldsymbol{\Lambda}_{\mathrm{GW}}) \pi(\boldsymbol{\Lambda}_{\mathrm{GW}}),
\end{align}
where $\pi(\boldsymbol{\Lambda}_{\mathrm{GW}})$ is the prior on the hyperparameters, and the likelihood is given by multiplying the individual-event likelihoods marginalized over the binary parameters $\boldsymbol{\theta}$~\citep{Thrane:2018qnx},
\begin{align}
    p(\{d\} | \boldsymbol{\Lambda}_{\mathrm{GW}}) \propto \frac{1}{\alpha(\boldsymbol{\Lambda}_{\mathrm{GW}})^{N}} \prod_{i}^{N}\sum_{j}\frac{\pi_{\mathrm{pop}}( \boldsymbol{\theta}_{i,j} | \boldsymbol{\Lambda}_{\mathrm{GW}})}{\pi_{\mathrm{PE}}(\boldsymbol{\theta}_{i,j})}
    \label{eq:hyper-like}.
\end{align}
Here, $\pi_{\mathrm{PE}}(\boldsymbol{\theta})$ is the original prior applied for the binary parameters $\boldsymbol{\theta}$ during the individual-event parameter estimation step, while $\pi_{\mathrm{pop}}(\boldsymbol{\theta} | \boldsymbol{\Lambda}_{\mathrm{GW}})$ is the population-level distribution we assume describes the data characterized by hyperparameters $\boldsymbol{\Lambda}_{\mathrm{GW}}$. The index $j$ represents the individual-event posterior samples, which we use to perform a Monte Carlo integral to marginalize over $\boldsymbol{\theta}$, and the index $i$ indicates the event in our set of four NSBHs. 

\begin{table}
	\centering
	\caption{Event names, FAR values, and references for the candidate signals included in our analysis.}
	\label{tab:events}
	\begin{tabular}{lcc} %
		\hline
		Name & FAR ($\mathrm{yr}^{-1}$) & Reference\\
		\hline
		GW190426\_152155 & $9.12\times 10^{-1}$ & \cite{LIGOScientific:2021usb}\\
		GW190917\_114630 & $6.56\times10^{-1}$ & \cite{LIGOScientific:2021djp}\\
  GW200105\_162426 & $2.04\times 10^{-1}$ & \cite{LIGOScientific:2021qlt}\\
		GW200115\_042309 & $< 10^{-5}$ & \cite{LIGOScientific:2021qlt}\\
		\hline
	\end{tabular}
\end{table}

For the individual-event posterior samples, we use the results publicly released by the LVK including the effects of gravitational-wave emission from higher-order modes and spin precession,\footnote{This corresponds to the \texttt{PrecessingSpinIMRHM} samples for the 2019 events~\citep{gwtc2_data_release_samples,gwtc2.1_data_release}, and the \texttt{C01:Mixed} samples for the 2020 events~\citep{gwtc3_data_release}.} where the prior on the dimensionless spin magnitude of both the black hole and neutron star covers the range $\chi \in [0,1]$ and the spin tilts can be misaligned relative to the orbital angular momentum. We model the neutron star as a point mass, as there are no currently-available waveform models that include the effects of higher order modes, misalignment of the tilt of the black hole spin, and the tidal deformability of the neutron star. However, previous works have found that the effect of tides on the waveform for NSBH sources is the least significant of the three aforementioned processes~\citep{Huang:2020pba}.

The term $\alpha(\boldsymbol{\Lambda}_{\mathrm{GW}})^{N}$ in the denominator of Eq.~\ref{eq:hyper-like} represents the fraction of sources drawn from a population model with hyperparameters $\boldsymbol{\Lambda}_{\mathrm{GW}}$ that would be detected. This correction accounts for the bias due to gravitational-wave selection effects in our chosen NSBH population and allows us to obtain unbiased estimates of the underlying astrophysical population, rather than the observed one~\citep{Loredo_2004, Mandel:2018mve, Vitale:2020aaz}. We evaluate $\alpha(\boldsymbol{\Lambda}_{\mathrm{GW}})$ by taking a Monte Carlo integral over detected events drawn from a simulated population following the method described in \cite{Farr_2019}. We use the sensitivity estimates for NSBH systems released at the end of the most recent LIGO-Virgo observing run (O3), obtained via a simulated injection campaign~\citep{ligo_scientific_collaboration_and_virgo_2021_5636816}.

We assume the black hole mass distribution follows a truncated power-law~\citep{Fishbach:2017zga},
\begin{align}
\label{eq:m1_pop}
    &\pi_{\mathrm{pop}}(m_{\mathrm{BH}} | \alpha, m_{\mathrm{BH}, \min}, m_{\mathrm{BH}, \max}) \propto \\ \nonumber
    &\qquad \begin{cases}
    m_{\mathrm{BH}}^{-\alpha}, &m_{\mathrm{BH}, \min} \leq m_{\mathrm{BH}} \leq m_{\mathrm{BH}, \max}\\
    0, &\mathrm{otherwise}
    \end{cases}.
\end{align}
We explore two different possibilities for the pairing function governing the distribution of the mass ratio between the black hole and the neutron star, $q \equiv m_{\mathrm{NS}}/m_{\mathrm{BH}}$: a truncated Gaussian or another power law~\citep{Fishbach:2019bbm},
\begin{align}
    \label{eq:q_pop_trunc}
    &\pi_{\mathrm{pop}}(q | m_{\mathrm{BH}}, m_{\mathrm{NS}, \max}, \mu, \sigma) \propto \\ \nonumber
    &\qquad \begin{cases}
    \mathcal{N}(q | \mu, \sigma), &q_{\min}(m_{\mathrm{BH}}) \leq q \leq q_{\max}(m_{\mathrm{BH}}, m_{\mathrm{NS}, \max})\\
    0, &\mathrm{otherwise}
    \end{cases},\\
    \label{eq:q_pop_pl}
    &\pi_{\mathrm{pop}}(q | m_{\mathrm{BH}}, m_{\mathrm{NS}, \max}, \beta) \propto \\ \nonumber
    &\qquad \begin{cases}
    q^\beta, &q_{\min}(m_{\mathrm{BH}}) \leq q \leq q_{\max}(m_{\mathrm{BH}}, m_{\mathrm{NS}, \max})\\
    0, &\mathrm{otherwise}
    \end{cases},
\end{align}
emphasizing that the pairing function is a conditional distribution that depends on the value of the black hole mass. Because we assume that the black hole is always the more massive (primary) compact object in the binary, so that $q \leq 1$, this means that the range of allowed mass ratio values changes depending on the black hole mass. We fix the minimum neutron star mass to $1~M_{\odot}$, so that $q_{\min} = 1/m_{\mathrm{BH}}$, and sample in the maximum neutron star mass, $m_{\mathrm{NS}, \max}$, as a free parameter, such that $q_{\max} = \min(m_{\mathrm{NS}, \max}/m_\mathrm{BH}, 1)$.

We fit the black hole spin with a Beta distribution with hyperparameters $\alpha_{\chi}$ and $\beta_{\chi}$~\citep{Wysocki:2018mpo},
\begin{align}
\label{eq:spin_pop}
    \pi_{\mathrm{pop}}(\chi_{\mathrm{BH}} | \alpha_{\chi}, \beta_{\chi}) = \frac{\chi_{\mathrm{BH}}^{\alpha-1}(1 - \chi_{\mathrm{BH}})^{\beta - 1}}{\mathrm{B}(\alpha_{\chi}, \beta_{\chi})},
\end{align}
where $\mathrm{B}(\alpha_{\chi}, \beta_{\chi})$ is the Beta function. We do not explicitly fit the spin of the neutron star, but restrict it to lie within the breakup spin, $\chi_{\mathrm{Kep}}$, which represents the maximum neutron star spin at the mass-shedding limit. The exact value of the breakup spin depends on the EoS, but is about $\chi_{\mathrm{Kep}}\sim0.7$ for most EoSs~\citep{Shao:2019ioq, Most:2020bba}.
We reweight the binary parameter posterior samples obtained under the precessing, high-spin prior for both the black hole and the neutron star so that the prior on the neutron star spin magnitude is uniform on $[0, 0.7]$.
Choosing a prior that supports high spins avoids biases that could arise in the mass distribution due to mismodeling the neutron star spin via the correlation between the component of the spin aligned with the orbital angular momentum and mass ratio~\citep{Biscoveanu:2021eht, Ye:2022qoe}.

Our full population prior, $\pi_{\mathrm{pop}}(\boldsymbol{\theta} | \boldsymbol{\Lambda}_{\mathrm{GW}})$, is the product of the population distributions for the black hole mass, mass ratio, and black hole spin given in Eqs.~\ref{eq:m1_pop}-\ref{eq:spin_pop}. We reweight the publicly-released posterior samples into a redshift prior that is uniform in comoving volume and source-frame time. We use the \textsc{dynesty} sampler~\citep{Speagle:2019ivv} as implemented in the \textsc{GWPopulation}~\citep{Talbot:2019okv} and \textsc{bilby}~\citep{Ashton:2018jfp} packages to draw samples from the posterior on $\boldsymbol{\Lambda}_{\mathrm{GW}}$ in Eq.~\ref{eq:hyper-post}. We apply uniform prior distributions for all the hyperparameters, $\pi(\boldsymbol{\Lambda}_{\mathrm{GW}})$, whose minimum and maximum values are given in Table~\ref{tab:priors}. We note that these prior ranges by definition exclude GW190814 from our target population, as its primary mass is greater than the maximum black hole mass of $20~M_{\odot}$ that we consider in our analysis.

\begin{table}
	\centering
	\caption{Hyperparameters describing the mass and spin distributions and the maximum and minimum values allowed in the prior applied during hierarchical inference. The priors on all parameters are uniform. }
	\label{tab:priors}
	\begin{tabular}{lcll} %
		\hline
		Symbol & Parameter & Minimum & Maximum\\
		\hline
		$\alpha$ & black hole mass power-law index & -4 & 12\\
		$m_{\mathrm{BH}, \min}$ & minimum black hole mass & $2~M_{\odot}$ & $10~M_{\odot}$\\
		$m_{\mathrm{BH}, \max}$ & maximum black hole mass & $8~M_{\odot}$ & $20~M_{\odot}$\\
		$m_{\mathrm{NS}, \max}$ & maximum neutron star mass & $1.97~M_{\odot}$ & $2.7~M_{\odot}$\\
		$\mu$ & mass ratio mean & 0.1 & 0.6\\
		$\sigma$ & mass ratio standard deviation & 0.1 & 1\\
		$\beta$ & mass ratio power-law index & -10 & 4\\
		$\alpha_{\chi}$ & black hole spin $\alpha$ & 0.1 & 10\\
		$\beta_{\chi}$ & black hole spin $\beta$ & 0.1 & 10\\
		\hline
	\end{tabular}
\end{table}

\subsection{Results}
\begin{figure}
	\includegraphics[width=\columnwidth]{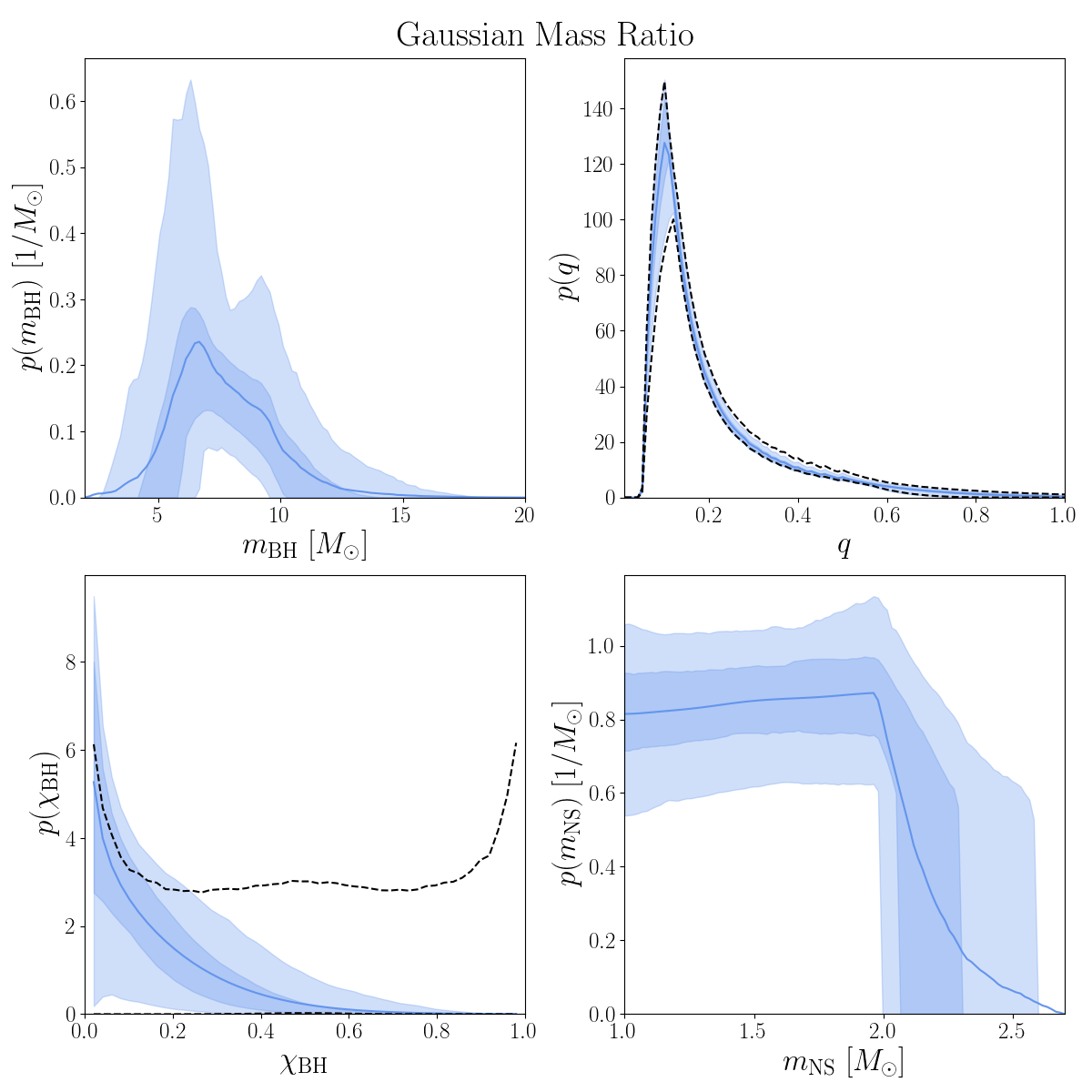}
	\includegraphics[width=\columnwidth]{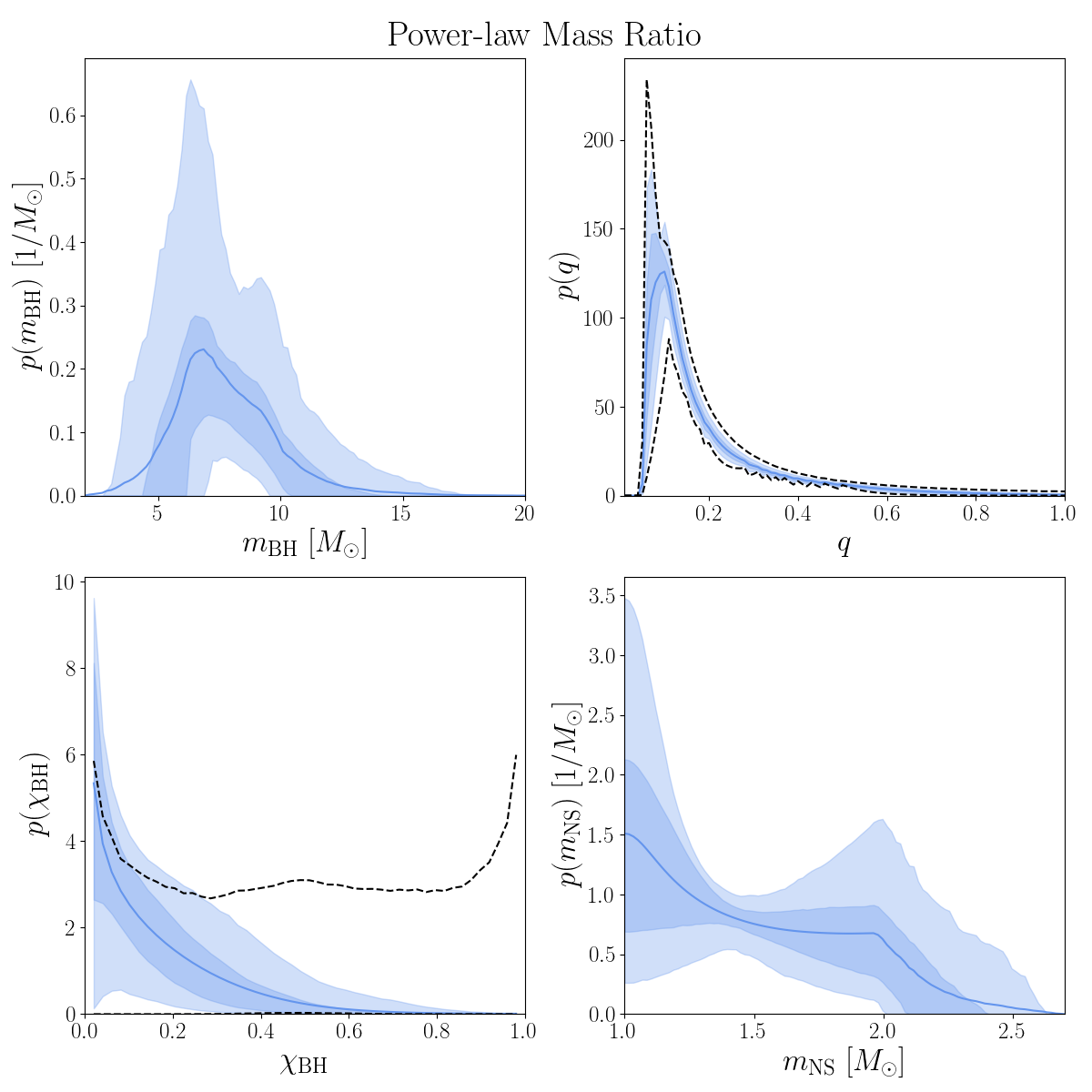}
    \caption{Posterior predictive distributions (solid blue) and 50\% and 90\% credible intervals (shaded blue) for the component masses, mass ratio, and black hole spin in the underlying, astrophysical NSBH population under the Gaussian mass ratio model (top) and power-law model (bottom). The black dashed lines show the 90\% credible region enclosed by draws from the hyperparameter prior for the black hole spin and mass ratio.}
    \label{fig:ppds}
\end{figure}

\begin{figure*}
	\includegraphics[width=\columnwidth]{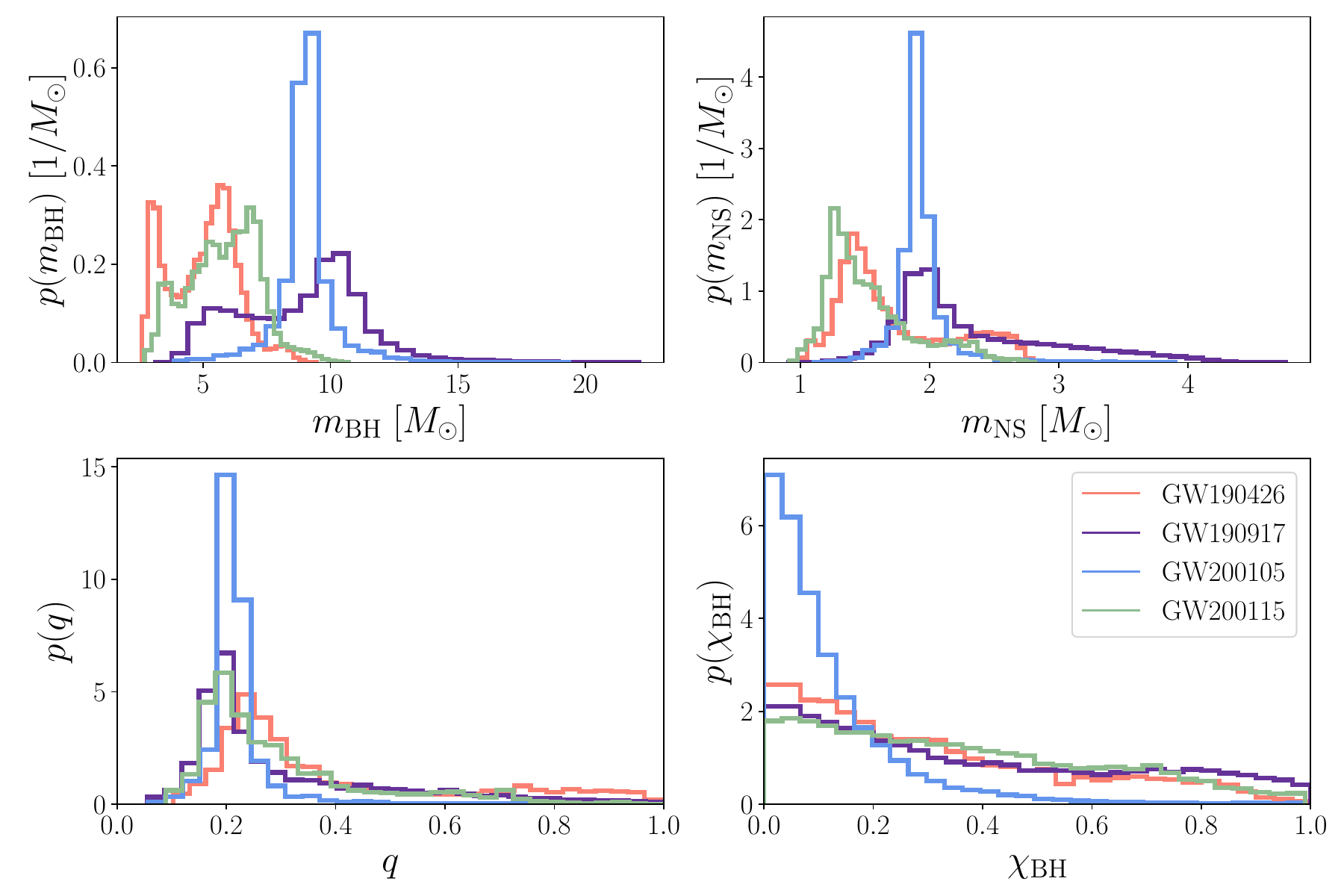}\qquad
	\includegraphics[width=\columnwidth]{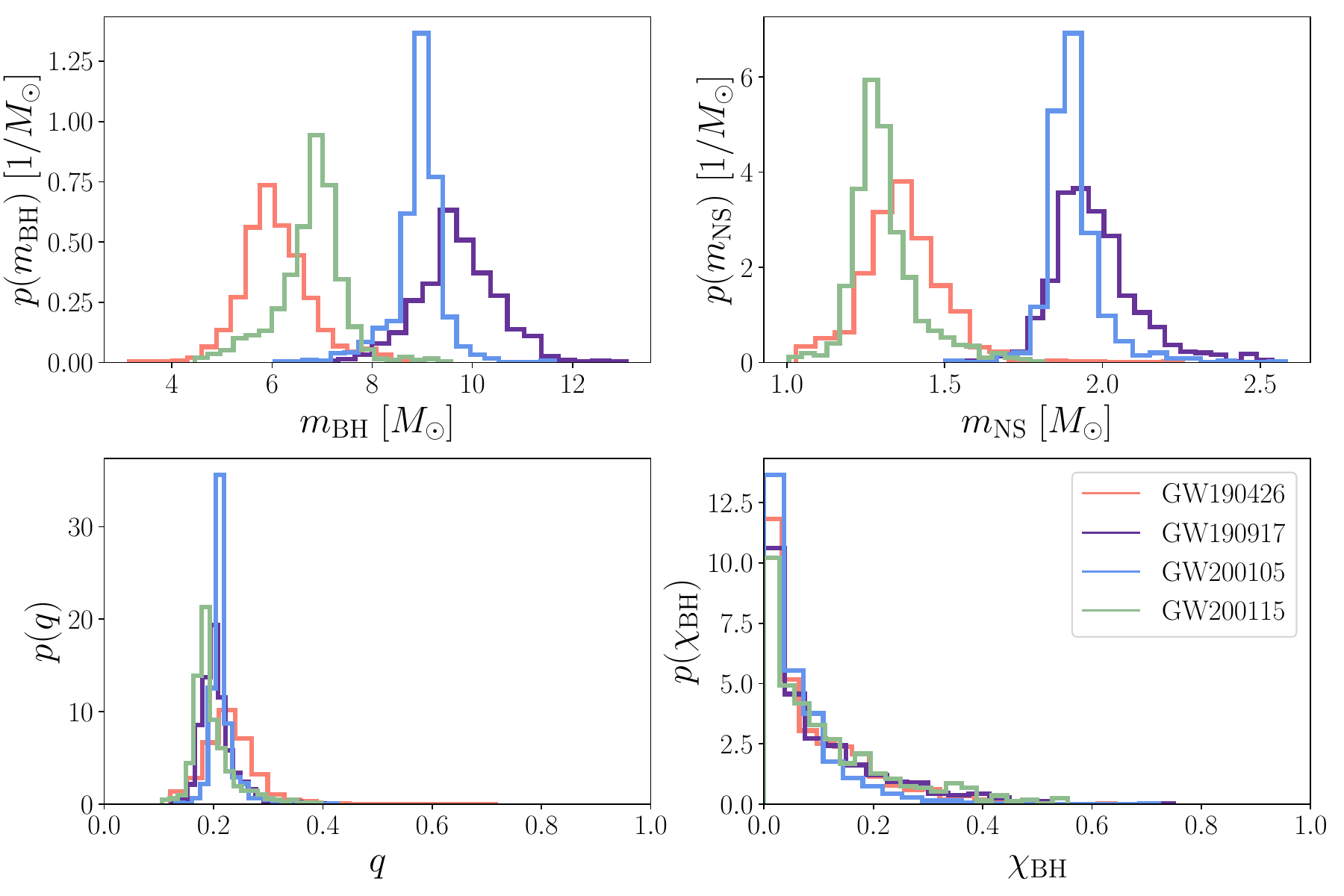}
    \caption{Distributions of the component masses, mass ratio, and black hole spin for each of the four individual events in our NSBH population under the original priors (left) and reweighted into the population prior inferred under the Gaussian mass ratio model (right).}
    \label{fig:1d_hists}
\end{figure*}

In Fig.~\ref{fig:ppds} we show the inferred component mass, mass ratio, and black hole spin posterior population distributions (PPDs), which are the expected distributions for the binary parameters, $\boldsymbol{\theta}$, of the \emph{astrophysical} population of new NSBH events irrespective of detectability inferred from the accumulated set of four detections:
\begin{align}
    \label{eq:ppd}
    p(\boldsymbol{\theta} | \{d\}) = \int \pi_{\mathrm{pop}}( \boldsymbol{\theta} | \boldsymbol{\Lambda}_{\mathrm{GW}}) p(\boldsymbol{\Lambda}_{\mathrm{GW}} | \{d\})d\boldsymbol{\Lambda}_{\mathrm{GW}}.
\end{align}
The results obtained under the Gaussian (top) and power-law (bottom) mass ratio models are qualitatively similar. The black hole mass distribution shown in the top left panels of each grid is constrained between $5.52^{+1.29}_{-2.66}-9.96^{+8.54}_{-1.05}~M_{\odot}$ (maximum posterior value on the minimum and maximum black hole masses and 90\% credible intervals calculated with the highest posterior density method) for the Gaussian pairing function, with discernible peaks at the maximum posterior values of both the minimum and maximum mass. This is due to a degeneracy between the maximum black hole mass, $m_{\mathrm{BH}, \max}$ and the black hole mass power-law index, $\alpha$. The posterior on $\alpha$ has support for both positive and negative slopes; the branch with support for positive slopes strongly prefers a peak at $m_{\mathrm{BH}, \max} \sim 10~M_{\odot}$, while the negative-slope branch supports a wide range of maximum black hole masses. This can be understood in terms of the black hole mass posteriors for the individual events, shown in Fig.~\ref{fig:1d_hists}, as the posterior support for all four events falls off at $m_{\mathrm{BH}}\gtrsim 10~M_{\odot}$. The maximum black hole mass inferred for the NSBH population using these four sources is significantly smaller than that inferred from the BBH population, which extends up to $80.60^{+18.68}_{-1.70}~M_{\odot}$~\citep{ligo_scientific_collaboration_and_virgo_2021_5636816},\footnote{We use the publicly-released hyperparameter samples obtained under the \textsc{Power Law + Peak} mass model and \textsc{Default} spin model presented in \cite{LIGOScientific:2021psn} for all statements about the BBH population throughout the manuscript.} indicating that the black holes in NSBH systems are systematically lighter than those in BBH systems.

Our results also indicate that the black holes in NSBHs have systematically smaller spins than those in BBHs. Under both mass ratio models, the inferred spin distribution shown in the bottom left panels of each grid falls off steeply from $\chi_{\mathrm{BH}}=0$ and only extends up to a maximum spin magnitude of $\chi_{\mathrm{BH}, 99}\leq 0.68$ under both the Gaussian and power-law pairing functions at 90\% credibility.\footnote{$\chi_{\mathrm{BH}, 99}$ represents the spin at which 99\% of the probability for the Beta distribution is enclosed.}
Support for high spin magnitudes is strongly suppressed relative to the 90\% credible region encompassed by samples from the hyperparameter prior, shown in the dashed black line.
Some of the differences between our inferred spin distribution and the BBH spin distribution presented in \cite{LIGOScientific:2021psn} can be attributed to differences in the prior on the spin hyperparameters (see Appendix~\ref{ap:bh_spin} for a detailed comparison), but the BBH spin distribution does not fall off as steeply and extends to higher spin magnitudes.

The upper right panels of each grid show the inferred mass ratio distribution under each mass ratio model. The shape of the mass ratio distribution is much more strongly determined by the priors on the maximum and minimum black hole and neutron star masses than by the functional form of the pairing function. Since the neutron star can only take on masses in the range $m_{\mathrm{NS}} \in [1, 2.7]~M_{\odot}$ and the black hole mass distribution covers the range $m_{\mathrm{BH}} \in [2, 20]~M_{\odot}$, this means that much of the prior probability is clustered around $q\sim0.1$, with a lower bound of $q=0.05$ set by the minimum neutron star and maximum black hole masses.
Particularly for the Gaussian model, there is very little information gained in the posterior shown in blue relative to the 90\% credible region allowed by the prior, shown in the dashed black line. 
We are only able to exclude narrow distributions peaked towards equal masses, with $\sigma\lesssim0.2$ and $\mu \gtrsim0.5$. Even for these distributions with higher values of $\mu$, the nature of the priors on the component mass ranges is such that the Gaussian is truncated well below the peak for all but the lowest black hole masses, which means that most of the probability still lies around $q\sim0.1$
The posterior on the power-law index under the power-law model is more informative, $\beta = -1.42^{+4.72}_{-3.32}$, although the shape of the mass ratio distribution under this model is similarly dominated by the choice of component mass ranges. 

The implied neutron star mass distributions given the inferred black hole mass distributions and pairing function are shown in the remaining panel of each grid. While the shapes of the distributions obtained under the two mass ratio models are different, the constraints on the maximum neutron star mass are similar, $m_{\mathrm{NS},\max}=2.07^{+0.59}_{-0.10}~M_{\odot}$ for the power-law model and $m_{\mathrm{NS},\max}=2.03^{+0.56}_{-0.06}~M_{\odot}$ for the Gaussian model. The posterior on $m_{\mathrm{NS}, \max}$ peaks at the lower bound of the prior, which is a conservative lower limit on the maximum NS mass observed electromagnetically~\citep{Antoniadis:2013pzd, NANOGrav:2019jur, Fonseca:2021wxt}. The posterior support does extend up to higher masses rather than railing narrowly against the lower edge of the prior, which would indicate that the true value of the maximum neutron star mass may actually lie below the lower edge of the prior. 
This would hint that the maximum mass of neutron stars in NSBHs detected in gravitational waves is smaller than the maximum mass of neutron stars detected electromagnetically, on which the value of the lower edge of the prior on $m_{\mathrm{NS},\max}$ is based. Instead, we find that the two are consistent.
The shape of the neutron star mass distribution is flatter under the Gaussian model, falling off sharply at the maximum mass, while for the power-law model the distribution peaks at the lower edge of the allowed mass range, but the posterior still supports a flat distribution. The effect of reweighting the individual-event posterior samples by the inferred population distribution, as shown on the right in Fig.~\ref{fig:1d_hists}, is to suppress the high-spin, equal-mass, and high-neutron-star-mass tails present under the original prior.

\begin{figure*}
	\includegraphics[width=0.95\columnwidth]{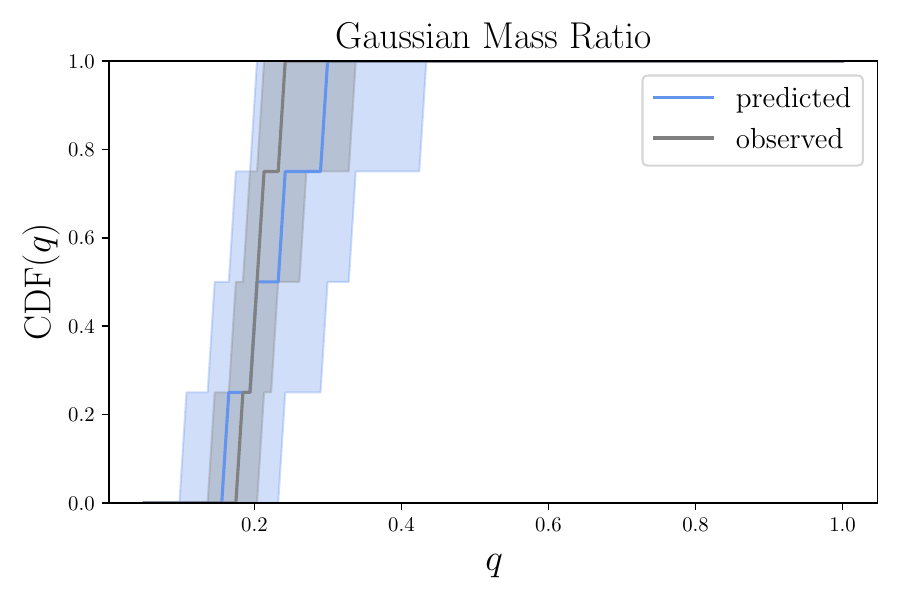}\qquad
	\includegraphics[width=0.95\columnwidth]{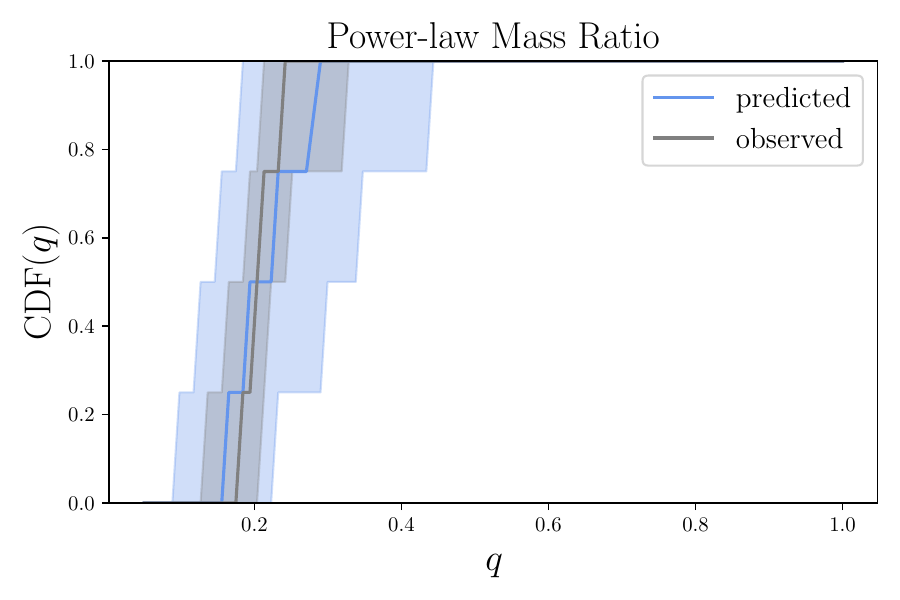}
    \caption{Posterior predictive check comparing the inferred population under the Gaussian (blue, left) and power-law (blue, right) mass ratio models to the observed population (grey).}
    \label{fig:ppcs}
\end{figure*}

In order to determine if there is a statistical preference between the Gaussian and power-law pairing functions, we perform a posterior predictive check to compare the inferred detectable populations under each model to the observed population. To do this, we reweight the simulated detected events used to compute $\alpha(\boldsymbol{\Lambda}_{\mathrm{GW}})$ in Eq.~\ref{eq:hyper-like} by the full population distribution $\pi_{\mathrm{pop}}(\boldsymbol{\theta} | \boldsymbol{\Lambda}_{\mathrm{GW}})$ implied by each hyperparameter posterior sample. For each hyperparameter posterior sample, we draw $N=4$ simulated events from the reweighted set of detected events. We then compute the CDF of the mass ratio of the four events, showing the median and 90\% credible interval of the CDFs across all hyperparameter posterior samples in blue in Fig.~\ref{fig:ppcs}. To compare against the observed population, we draw one sample from the posterior on the mass ratio reweighted into the population prior implied by each hyperparameter posterior sample for each of the four real detected events (one sample from each histogram in the right panel of Fig.~\ref{fig:1d_hists}). We compute the CDF of these four samples and show the median and 90\% credible interval in grey in Fig.~\ref{fig:ppcs}. 

The results for the Gaussian pairing function are shown on the left and the power-law pairing results are shown on the right. The discrete steps in the CDFs come from the fact that each is computed using only four samples. The CDF range for the observed events lies within the predicted CDF range based on the population results for both models. Because we have so few observations, there is considerable uncertainty in both the predicted and observed distributions, and the posterior predictive check does not lend more support to one model or the other. Both provide suitable fits to the data given the limited number of observations in our sample. This is consistent with the fact that the shapes of the mass ratio distributions for the Gaussian and power-law models are so similar due to their dependence on the component mass ranges via the minimum and maximum mass ratios. We note that while the observed CDFs peak around $q\sim 0.2$, the astrophysical mass ratio distribution shown in Fig.~\ref{fig:ppds} instead peaks at lower mass ratios, around $q\sim0.1$. At fixed chirp mass, binaries with more equal mass ratios take longer to merge and hence have more time to accumulate SNR in the detector, making them easier to detect. This selection effect explains the shift towards more equal mass ratios between the astrophysical and observed distributions.

\subsection{EM-bright fraction}
\label{sec:em-bright}
As an extension of the population measurements reported in the previous section, we can calculate the implied fraction of GW-detectable NSBH systems which can produce an electromagnetic counterpart due to tidal disruption of the neutron star outside the black hole's innermost stable circular orbit (ISCO) radius, $R_{\mathrm{isco}}$, before the merger. We use the fitting formula for the remnant mass $\hat{M}_{\mathrm{rem}}$ presented in \cite{Foucart:2018rjc}, which depends on the symmetric mass ratio, $\eta = q/(1+q)^{2}$, black hole spin aligned with the orbital angular momentum, $\chi_{\mathrm{BH}, z}$, and neutron star EoS via the compactness, $C_{\mathrm{NS}}=Gm_{\mathrm{NS}}/(R_{\mathrm{NS}}c^{2})$:
\begin{align}
    \hat{M}_{\mathrm{rem}} = m_{\mathrm{NS}, b}\left[\max\left( 
        \alpha \frac{(1 - 2 C_{\mathrm{NS}})}{\eta^{1/3}}
        - \beta \frac{R_{\mathrm{isco}}(\chi_{\mathrm{BH},z}) C_\mathrm{NS}}{\eta} 
        + \gamma, 0\right)\right]^{\delta}.
        \label{eq:mrem}
\end{align}
Here, $m_{\mathrm{NS},b}$ is the baryonic mass of the neutron star (the sum of its binding energy and gravitational mass $m_{\mathrm{NS}}$), and $\alpha = 0.406, \beta = 0.139, \gamma = 0.255, \delta = 1.761$. Although this fitting formula was developed for NSBH mergers with nonrotating NSs, in the absence of a more general fit in the literature, we assume that it is also valid for spinning NSs if we take $C_{\rm NS}$ to be the compactness of the rotating NS.

To express the baryonic mass as an approximate function of the gravitational mass and spin for any equation of state, we use the empirical fitting formula from \cite{Cipolletta:2015nga}:
\begin{align}
\label{eq:baryonic_mass}
    \frac{m_{\mathrm{NS}, b}}{M_{\odot}} = \frac{m_{\mathrm{NS}}}{M_{\odot}} + \frac{13}{20}\left(\frac{m_\mathrm{NS}}{M_{\odot}}\right)^2\left(1 - \frac{1}{130}\left(\frac{m_\mathrm{NS}}{M_{\odot}}\right)^{3.4}{\chi_{\mathrm{NS}}}^{1.7}\right).
\end{align}
While the baryonic mass itself does not depend on the neutron star spin, the relationship between the baryonic and gravitational masses does, since the properties of spinning neutron stars differ from those of their non-spinning counterparts. Spinning neutron stars are supported against gravitational collapse up to higher masses due to their rotation. We follow \citet{Ye:2022qoe} and calculate the maximum neutron star mass accounting for rigid rotation using the universal relation presented in \citet{Breu:2016ufb, Most:2020bba},
 \begin{align}
    \label{eq:spin}
     m_{\mathrm{NS, crit}} = m_{\mathrm{TOV}}\left[1 + a_1 \left(\frac{\chi_{\mathrm{NS}}}{\chi_{\mathrm{Kep}}}\right)^{2} + a_2 \left(\frac{\chi_{\mathrm{NS}}}{\chi_{\mathrm{Kep}}}\right)^{4}\right],
 \end{align}
 with $a_1 = 0.132,\ a_2 = 0.071$, and $m_{\mathrm{TOV}}$ is the maximum mass that can be supported against gravitational collapse for a non-spinning neutron star. The breakup spin introduced in the previous section, $\chi_{\mathrm{Kep}}$, can be expressed in terms of the compactness at the TOV mass, $C_{\mathrm{TOV}} = C_{\mathrm{NS}}(m_{\mathrm{TOV}})$, as
 \begin{align}
    \chi_{\mathrm{Kep}} = \frac{\alpha_1}{\sqrt{C_{\mathrm{TOV}}}} + \alpha_2\sqrt{C_{\mathrm{TOV}}},
 \end{align}
with $\alpha_1 = 0.045,\ \alpha_2 = 1.112$~\citep{Breu:2016ufb, Koliogiannis:2019rvh, Most:2020kyx, Shao:2019ioq}. 

The gravitational mass of a spinning neutron star is higher than the gravitational mass of the NS with the same central density at rest, and its equatorial radius is also larger. However, \citet{Konstantinou:2022vkr} find that the compactness of a rotating neutron star is the same as the compactness of the non-rotating neutron star with the same central density to within a few percent for astrophysically-realistic values of the compactness up to the breakup spin. If we can compute the radius of the non-spinning neutron star with the same central density, this allows us to calculate $C_{\mathrm{NS}} \approx C_{\mathrm{NS, 0}}$. Because the gravitational mass is the quantity that we measure with gravitational waves but the mass-radius relation for each EoS is given in terms of the non-spinning mass and radius, we need a way to map from the gravitational mass of the spinning neutron star to its rest mass. We assume that the expression in Eq.~\ref{eq:spin} holds for all neutron star masses, such that 
\begin{align}
    \label{eq:m_nonspinning}
     m_{\mathrm{NS},0} = m_{\mathrm{NS}}\left[1 + a_1 \left(\frac{\chi_{\mathrm{NS}}}{\chi_{\mathrm{Kep}}}\right)^{2} + a_2 \left(\frac{\chi_{\mathrm{NS}}}{\chi_{\mathrm{Kep}}}\right)^{4}\right]^{-1},
 \end{align}
where we use $m_{\mathrm{NS}}$ to indicate the neutron star mass measured with gravitational waves and $m_{\mathrm{NS},0}$ to indicate the rest mass of the object with measured gravitational mass $m_{\mathrm{NS}}$ and spin $\chi_{\mathrm{NS}}$.\footnote{We opt to use the approximation in Eq.~\ref{eq:m_nonspinning} rather than the universal relation between the spinning and non-spinning gravitational masses presented in \cite{Konstantinou:2022vkr}, as the latter requires calculating the EoS-dependent moment of inertia to convert between dimensionless spin and angular rotation frequency.} The various neutron star masses that we use in our analysis are summarized in Table~\ref{tab:ns_masses}.

\begin{table*}
	\centering
	\caption{Descriptions of the various neutron star mass parameters used in our analysis.}
	\label{tab:ns_masses}
	\begin{tabular}{lcl} %
		\hline
		Symbol & Definition & Equation \\
		\hline
		$m_{\mathrm{NS}}$ & gravitational mass including the effect of spin & - \\
		$m_{\mathrm{NS},b}$ & baryonic mass, binding energy + gravitational mass & \ref{eq:baryonic_mass} \\
		$m_{\mathrm{NS}, 0}$ & gravitational mass of the non-spinning neutron star with $m_{\mathrm{NS}}, \chi_{\mathrm{NS}}$ & \ref{eq:m_nonspinning}\\
		$m_{\mathrm{TOV}}$ & maximum mass of a non-spinning neutron star for a particular EoS & - \\
		$m_{\mathrm{NS, crit}}$ & maximum mass of a spinning neutron star for a particular EoS & \ref{eq:spin}\\
		\hline
	\end{tabular}
\end{table*}

To account for the uncertainty in the EoS in our calculation of $\hat{M}_{\mathrm{rem}}$, we marginalize over the publicly released non-parametric EoS posterior samples~\citep{legred_isaac_2022_6502467} conditioned on data from gravitational-wave observations of binary neutron star mergers and radio and X-ray pulsar observations obtained by \cite{Legred:2021hdx}. We associate each mass and spin population hyperparameter posterior sample with an EoS sample, $\boldsymbol{\Lambda}_{\mathrm{EoS}}$, requiring that the critical mass in Eq.~\ref{eq:spin} for that EoS is greater than or equal to the value of $m_{\mathrm{NS},\max}$ for that hyperparameter sample, so that the maximum mass in the astrophysical population is within the maximum mass supported by that EoS. 

 For each sample from $\boldsymbol{\Lambda} = (\boldsymbol{\Lambda}_{\mathrm{GW}}, \boldsymbol{\Lambda}_{\mathrm{EoS}})$, we reweight the simulated detected events used to calculate $\alpha(\boldsymbol{\Lambda}_{\mathrm{GW}})$ in Eq.~\ref{eq:hyper-like} by the implied population distribution. For the neutron star in each of the reweighted NSBH binaries, we calculate the radius and compactness given by the EoS defined by that $\boldsymbol{\Lambda}$ sample by interpolating the non-spinning mass-radius relation given in the \cite{legred_isaac_2022_6502467} dataset. 
 If a particular neutron star sample is above the maximum neutron star mass supported by that EoS, $m_{\mathrm{NS, crit}}$, we remove that sample from our population, which is by definition restricted only to valid NSBH systems. 

 With the compactness calculated for each neutron star, we then calculate $\hat{M}_{\mathrm{rem}}$ for each of the binaries in the reweighted population. Because there is no universally accepted threshold for the remnant mass above which a NSBH merger will be electromagnetically bright (EM-bright)\footnote{Precursor emission produced before the NSBH merger has also been studied as a potential electromagnetic counterpart, sourced either via magnetospheric interactions or neutron star tidal resonances~\citep[see, e.g.,][]{Fernandez:2016sbf}. We do not consider the possibility of such precursor emission in the formulation of our analysis.}, we report our results for the EM-bright fraction in terms of the fraction of GW-detectable sources for each $\boldsymbol{\Lambda}$ sample for which $\hat{M}_{\mathrm{rem}} \geq M_{\mathrm{rem, min}}$, with $M_{\mathrm{rem, min}} = 0, 10^{-2}, 10^{-1}~M_{\odot}$.
 By reweighting the simulated detected events, we are quantifying $f_{\text{EM-bright}}$ for the population of NSBH mergers detectable in gravitational waves rather than the underlying astrophysical population. 
 
\begin{figure}
	\includegraphics[width=\columnwidth]{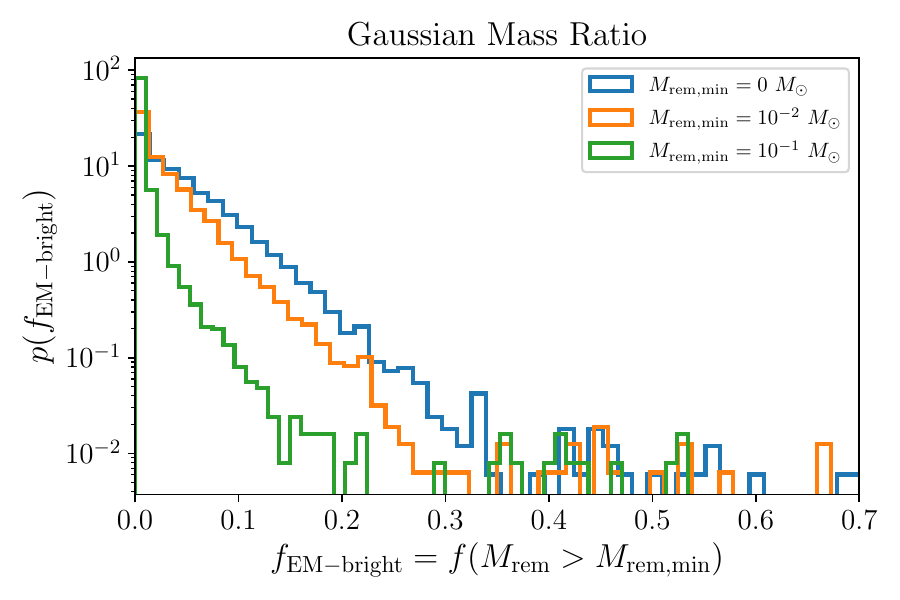}
	\includegraphics[width=\columnwidth]{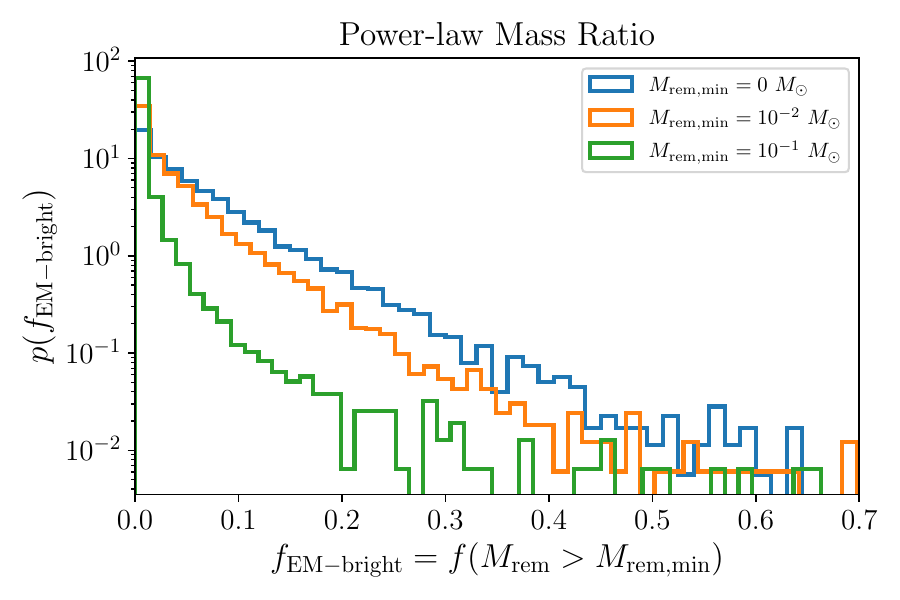}
    \caption{Posterior on the fraction of GW-detectable NSBH systems that will be electromagnetically bright with remnant mass $\hat{M}_{\mathrm{rem}} \geq M_{\mathrm{rem, min}}$, with the different colors indicating different values of $M_{\mathrm{rem, min}}$. The posterior is marginalized over the uncertainty in the neutron star equation of state and in the population hyperparameters for both the Gaussian (top) and power-law (bottom) pairing functions.}
    \label{fig:em_bright}
\end{figure}

 The posteriors for $f_{\text{EM-bright}} = f(\hat{M}_{\mathrm{rem}} \geq M_{\mathrm{rem, min}})$ shown in Fig.~\ref{fig:em_bright} are strongly peaked at $f_{\text{EM-bright}}=0$ with $f(\hat{M}_{\mathrm{rem}} \geq 0~M_{\odot})\leq 0.11$ at 90\% credibility for the Gaussian pairing function and $f(\hat{M}_{\mathrm{rem}} \geq 0~M_{\odot})\leq 0.14$ for the power-law pairing. This means that at most 14\% of GW-detectable sources will have a chance to be EM-bright.  The posteriors on the EM-bright fraction using $M_{\mathrm{rem, min}}=10^{-4}, 10^{-3}~M_{\odot}$ are statistically similar to the result obtained with $M_{\mathrm{rem, min}}=0$, leading us to conclude that we cannot distinguish the effect of values of the minimum remnant mass below $10^{-3}$ on the NSBH EM-bright fraction. The results under the power-law pairing function extend to larger values of $f_{\mathrm{EM-bright}}$, which can be explained in terms of the more gradual drop-off from the peak of the mass ratio distribution relative to the Gaussian pairing shown in Fig.~\ref{fig:ppds}. More equal mass ratios lead to more significant neutron star tidal disruption and a larger remnant mass. A similar effect is responsible for the samples predicting the largest EM-bright fractions, as those correspond to the hyperparameter posteriors that favor black hole mass distributions that peak at small masses with a large negative slope, leading to more equal mass ratios.
 
 We can use the same method to place constraints on the absolute contribution of NSBHs to heavy metal production in the universe via the ejection of r-process elements. Instead of sampling from the detectable population of NSBH mergers as we did for the calculation of $f_{\text{EM-bright}}$, we sample from the inferred underlying astrophysical population of sources (shown in Fig.~\ref{fig:ppds}) in proportion to our inferred rate of NSBH mergers, marginalizing over the uncertainty in the population properties, merger rate, and EoS using the \cite{Legred:2021hdx} posterior samples. We obtain a merger rate in the range $1.24-62.3~\mathrm{Gpc^{-3}yr^{-1}}$ ($1.15-57.2~\mathrm{Gpc^{-3}yr^{-1}}$) under the Gaussian (power-law) pairing function assuming a prior $\pi(\mathcal{R})\propto 1/\mathcal{R}$ on the merger rate. Our corresponding posterior on the total r-process ejecta mass contribution from the astrophysical NSBH population rails strongly at $0~M_{\odot}$, with 90\% of the probability lying within $\hat{M}_{\mathrm{rem}}^{\mathrm{tot}} \leq 0.72~M_{\odot}/\mathrm{Gpc^{-3}yr^{-1}}$ ($\hat{M}_{\mathrm{rem}}^{\mathrm{tot}} \leq 1.03~M_{\odot}/\mathrm{Gpc^{-3}yr^{-1}}$). 
 
 To put this number in context, GW170817 produced at least 0.01 $M_\odot$ of ejecta (see, e.g., Table 1 of \citealt{Cote:2017evr} and references therein). Assuming this is typical for BNS mergers, and adopting the preferred GWTC-3 BNS rate estimate of 170 Gpc$^{-3}$ yr$^{-1}$~\citep{LIGOScientific:2021psn}, the total BNS ejecta mass yield is $\sim 1.70~M_{\odot}/\mathrm{Gpc^{-3}yr^{-1}}$. Thus, this back-of-the envelope estimate indicates that NSBHs contribute at most $\sim40\%$ of the r-process ejecta in the universe, consistent with the simulation-based estimates of \cite{Chen:2021fro}, although we emphasize that the exact fraction depends sensitively on the assumed BNS population properties.

\section{Multimessenger constraints on the EoS and NSBH population properties}
\label{sec:joint_analysis}
\subsection{Methods}
\label{sec:joint_methods}
We find that our population hyperparameter posteriors obtained using gravitational-wave data alone under the Gaussian (power-law) pairing function predict that there is a $98.8^{+1.2}_{-33.5}$\% ($98.8^{+1.2}_{-43.6}$\%) probability that none of the four detected NSBH systems were EM-bright. This probability is obtained by applying a binomial distribution to each of the posterior samples in $f_{\mathrm{EM-bright}}$ shown in Fig.~\ref{fig:em_bright} for $M_{\mathrm{rem, min}}=0~M_{\odot}$ with $p(\mathrm{success}) = f_{\mathrm{EM-bright}}$ for zero successes and four failures.
Using this bound, we seek to determine if a meaningful multimessenger constraint can be placed on the NSBH population properties and neutron star EoS using the lack of detection of any electromagnetic counterpart for our four observed sources. We make the assumption that no counterpart was detected for any of the observed gravitational-wave events because there was no counterpart to detect, namely $M_{\mathrm{rem}}=0$ for all four of the detected sources. We use $\hat{M}_{\mathrm{rem}}$ to indicate the predicted remnant mass given the NSBH binary parameters and EoS using Eq.~\ref{eq:mrem}, and $M_{\mathrm{rem}}$ to indicate a measured value of the remnant mass for an observed system. 

We emphasize that the assumption we make represents the best-case scenario for EoS constraints from nondetection; in reality a number of factors contribute to the detectability of the EM counterpart, such as the intrinsic brightness of the emission if $M_{\mathrm{rem}}>0$ depending on the lightcurve model, the sensitivity and coverage of the available telescope network, and the promptness and accuracy of the sky localization of the event released by the LVK~\citep[see, e.g.][]{Metzger:2011bv, Feeney:2020kxk}. We do not attempt to carefully account for these electromagnetic selection effects, as their contributions to the lack of counterpart detection are highly uncertain, and instead make the maximally optimistic assumption that $M_{\mathrm{rem}}=0$ for each of the four NSBH systems detected in gravitational waves: i.e., we assume that the EM surveys would have detected any $M_{\mathrm{rem}} > 0$. As such, the results we present based on this assumption represent a conservative upper limit on the constraining power of nondetection in the multimessenger analysis of NSBH sources.

In addition to the gravitational-wave data for each event, we now also include the measurement of the remnant mass of each event as an independent data point, assuming $M_{\mathrm{rem}}=0$. The incorporation of the remnant mass data into the hierarchical inference method amounts to the introduction of an additional term in the numerator of Eq.~\ref{eq:hyper-like}---the likelihood of observing remnant mass $M_{\mathrm{rem}}=0$ given the parameters $q, m_\mathrm{NS}, \chi_{\mathrm{NS}}, \chi_{\mathrm{BH}, z}$ and $\boldsymbol{\Lambda}_{\mathrm{EoS}}$,
\begin{align}
    &p(M_{\mathrm{rem}} | q, m_\mathrm{NS}, \chi_{\mathrm{BH}, z}, \boldsymbol{\Lambda}_{\mathrm{EoS}}) \\ &\qquad \qquad =\delta(M_{\mathrm{rem}} - \hat{M}_{\mathrm{rem}}(q, m_\mathrm{NS}, \chi_{\mathrm{NS}}, \chi_{\mathrm{BH}, z}, \boldsymbol{\Lambda}_{\mathrm{EoS}}))\\
    &\qquad \qquad = \delta(\hat{M}_{\mathrm{rem}}(q, m_\mathrm{NS}, \chi_{\mathrm{NS}}, \chi_{\mathrm{BH}, z}, \boldsymbol{\Lambda}_{\mathrm{EoS}})).
\end{align}
The full derivation of the multimessenger likelihood is given in Appendix~\ref{ap:math}. In practice, this means that in addition to the mass and spin hyperparameters, $\boldsymbol{\Lambda}_{\mathrm{GW}}$, we also now sample in a set of hyperparameters describing the EoS, which enter the likelihood via the compactness that goes into the calculation of $\hat{M}_{\mathrm{rem}}$. 

For the multimessenger analysis, we adopt the piecewise polytrope EoS characterized by four parameters introduced in \citet{Read:2008iy}, where the pressure as a function of density is given by
\begin{align}
    p(\rho) = K_{i}\rho^{\Gamma_{i}}
\end{align}
in each of three different regions, $i=1,2,3$, with transition densities of $\rho_{1} = 10^{14.7}~\mathrm{g/cm^3}$ and $\rho_{2} = 10^{15}~\mathrm{g/cm^3}$. Requiring that the pressure be continuous across the transition densities means the full EoS can be determined by four parameters---the three $\Gamma_{i}$ adiabatic indices and an overall pressure scale, $p_{1} = p(\rho_{1})$. Instead of using the \cite{legred_isaac_2022_6502467} posteriors on the EoS as was done in Section~\ref{sec:em-bright}, we now directly sample in $\boldsymbol{\Lambda}_{\mathrm{EoS}} = (\log(p_{1}/[\mathrm{dyne/cm^{2}}]), \Gamma_{1}, \Gamma_{2}, \Gamma_{3})$ to obtain an independent posterior from NSBH observations alone, applying a uniform prior on these EoS hyperparameters.

 We follow \cite{HernandezVivanco:2019vvk} in setting the prior ranges on $\boldsymbol{\Lambda}_{\mathrm{EoS}}$, shown in Table~\ref{tab:eos}, and in imposing two additional constraints. We require the maximum mass of a non-spinning neutron star supported by the EoS determined by a particular draw from $\boldsymbol{\Lambda}_{\mathrm{EoS}}$ to be $m_{\mathrm{TOV}} \geq 1.97~M_{\odot}$. This value is chosen to match the lower bound of the prior on the maximum neutron star mass hyperparameter described in Section~\ref{sec:gw_methods}, $m_{\mathrm{NS}, \max}$. We also require that the EoS does not violate causality, so that the speed of sound in the neutron star is less than the speed of light. Because of the accuracy limitations of the piecewise polytrope fit to the EoS, in practice we only enforce the causality constraint when the speed of sound exceeds $1.1c$. We calculate the radius and compactness given by the EoS defined by each $\boldsymbol{\Lambda}_{\mathrm{EoS}}$ sample using the \textsc{eosinference} package~\citep{eosinference, Lackey:2014fwa} in conjunction with Eq.~\ref{eq:m_nonspinning} to take into account the neutron star spin. 
 
  \begin{table}
	\centering
	\caption{Hyperparameters describing the piecewise polytrope neutron star equation of state and the maximum and minimum values allowed in the prior. The priors on all parameters are uniform.}
	\label{tab:eos}
	\begin{tabular}{lcll} %
		\hline
		Symbol & Parameter & Minimum & Maximum\\
		\hline
		$\log(p_{1}/[\mathrm{dyne/cm^{2}}])$ & log-pressure at $\rho_{1}$ & 33.6 & 34.8\\
		$\Gamma_{1}$ & first adiabatic index & 2 & 4.5\\
		$\Gamma_{2}$ & second adiabatic index & 1.1 & 4.5\\
		$\Gamma_{3}$ & third adiabatic index & 1.1 & 4.5\\
		\hline
	\end{tabular}
\end{table}
 
 We do not impose additional EoS prior information, e.g.,~from the neutron star tidal deformability derived from gravitational-wave observations of the binary neutron star mergers GW170817~\citep{LIGOScientific:2017vwq} and GW190425~\citep{LIGOScientific:2020aai}, so our prior, shown in the dashed black line in Fig.~\ref{fig:mass-radius_joint}, includes very stiff equations of state.\footnote{Our prior and model choice for the EoS parameterization is different than that used in \citet{Legred:2021hdx}, although we do not expect the model differences to qualitatively affect the comparison of the two results in Fig.~\ref{fig:mass-radius_joint}.} We make this choice to obtain an independent posterior on the EoS parameters using the NSBH data alone in order to compare the constraining power of this multimessenger analysis against the tidal information that can be extracted from GW observations of BNS mergers without any multimessenger observations.
 
For the multimessenger analysis, we only use the Gaussian pairing function model, since the gravitational-wave-only analysis revealed that the two pairing functions give statistically similar results. We no longer sample directly in $m_{\mathrm{NS}, \max}$, and instead use the spin-dependent critical neutron star mass calculated using Eq.~\ref{eq:spin} for each sample from $\boldsymbol{\Lambda}$, i.e.,~we impose that the maximum mass in the astrophysical NS population is the spinning maximum mass supported by the EoS, $m_{\mathrm{NS},\max}=m_{\mathrm{NS, crit}}$. This means that the maximum mass ratio, $q_{\max}$, defined in Eq~\ref{eq:q_pop_trunc} is now a function of $(m_{\mathrm{BH}}, \chi_{\mathrm{NS}}, \boldsymbol{\Lambda}_{\mathrm{EoS}})$ rather than just $m_{\mathrm{BH}}$ and the single hyperparameter $m_{\mathrm{NS},\max}$. We also impose that the neutron star spin should lie within the specific value of the breakup spin, $\chi_{\mathrm{Kep}}$, defined by each EoS sample rather than setting a universal value of $\chi_{\mathrm{Kep}}=0.7$ as was done in the GW-only analysis. For this multimessenger analysis, we generate samples from the posterior on $\boldsymbol{\Lambda}$ using the \textsc{nestle} sampler~\citep{nestle}.

Since we allow the NS and BH mass ranges to overlap in order to probe the presence of a mass gap between them (see prior ranges in Table~\ref{tab:priors}), it is possible that a binary in our population with two equal-mass components of $2~M_{\odot}$ each, for example, could also be either a BNS or BBH by our own definitions of the NS and BH mass ranges. By insisting that all four of our events are NSBHs, we are imposing prior knowledge of source classification, which can lead to an overestimation of the information provided by a nondetection. If the population actually includes some low-mass BBHs, they should be discarded since they do not contribute any information to the constraint on the NS EoS. For now we include the simplifying assumption that all sources in the population are NSBHs, in the spirit of being maximally optimistic about the constraining power of our multimessenger analysis.

\subsection{Results}
\label{sec:joint_results}
\begin{figure}
	\includegraphics[width=\columnwidth]{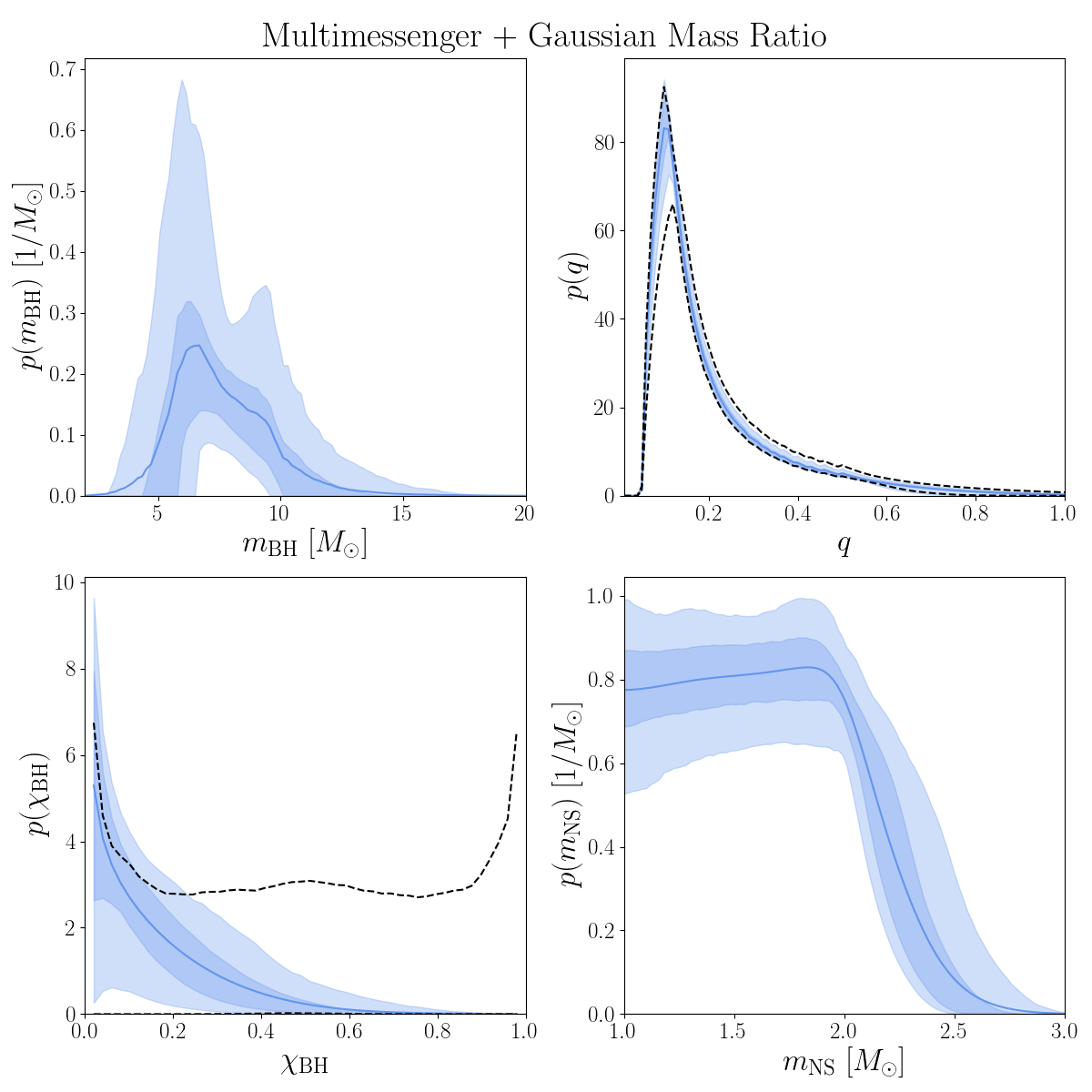}
    \caption{Posterior predictive distributions (solid blue) and 50\% and 90\% credible intervals (shaded blue) for the component masses, mass ratio, and black hole spin in the underlying, astrophysical population under the Gaussian mass ratio model from the multimessenger analysis that assumes none of the four detected NSBHs produced ejecta. The black dashed lines show the 90\% credible region enclosed by draws from the hyperparameter prior for the black hole spin and mass ratio.}
    \label{fig:ppds_joint}
\end{figure}

The population distributions inferred for the NSBH binary parameters folding in the nondetection of any electromagnetic counterparts are shown in Fig.~\ref{fig:ppds_joint}. The results are very similar to those obtained using the GW data only, shown in the top panel of Fig.~\ref{fig:ppds}, particularly for the black hole mass and spin distributions. The implied distribution for the neutron star mass tapers off more gradually as the upper limit is EoS-dependent under the multimessenger model, unlike the universal, hard cutoff at $m_{\mathrm{NS}, \max}$ under the GW-only model.  Because we are assuming $m_{\mathrm{NS}, \max}=m_{\mathrm{NS, crit}}$ in the multimessenger case, the maximum neutron star mass depends on the spin and the EoS following Eq.~\ref{eq:spin}. The posterior on the mass ratio distribution slightly prefers more extreme mass ratios for the multimessenger analysis compared to the GW-only analysis. This extra information on the mass ratio comes from the effect of $q$ on the remnant mass. More equal mass ratios lead to more remnant mass, so enforcing that $M_{\mathrm{rem}}=0$ pushes the posterior on the mean of the mass ratio distribution, $\mu$, towards lower values. 

\begin{figure*}
	\centering
	\includegraphics[width=0.8\columnwidth]{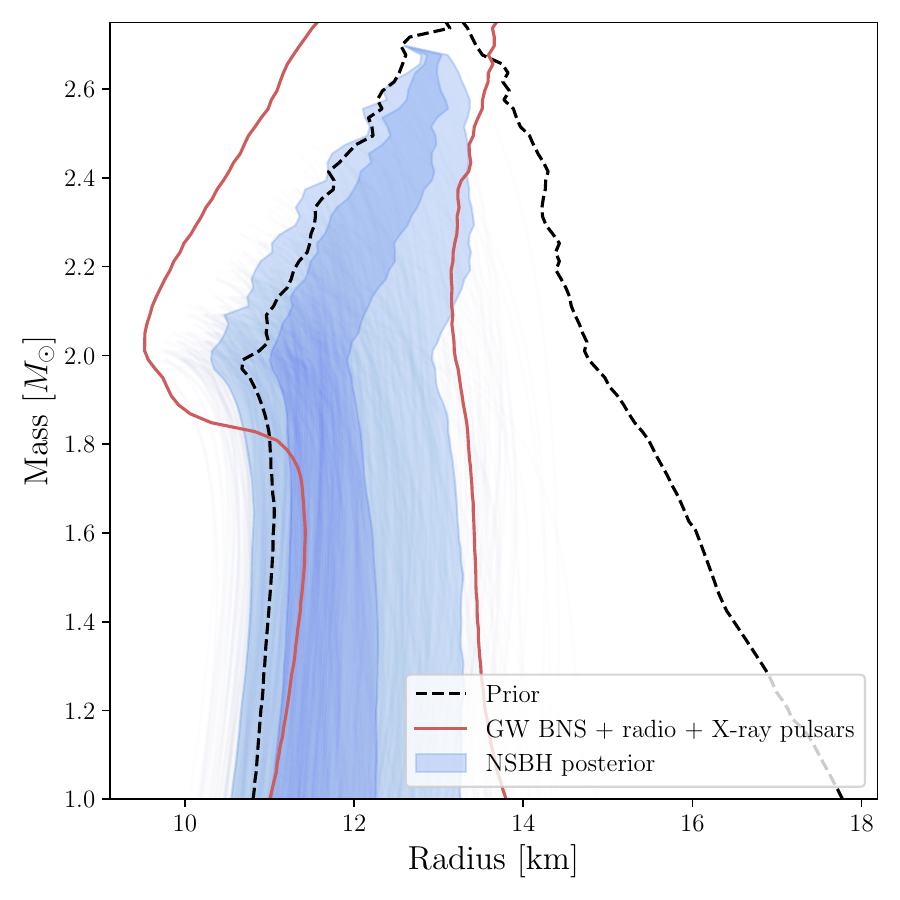}
	\qquad \includegraphics[width=0.89\columnwidth]{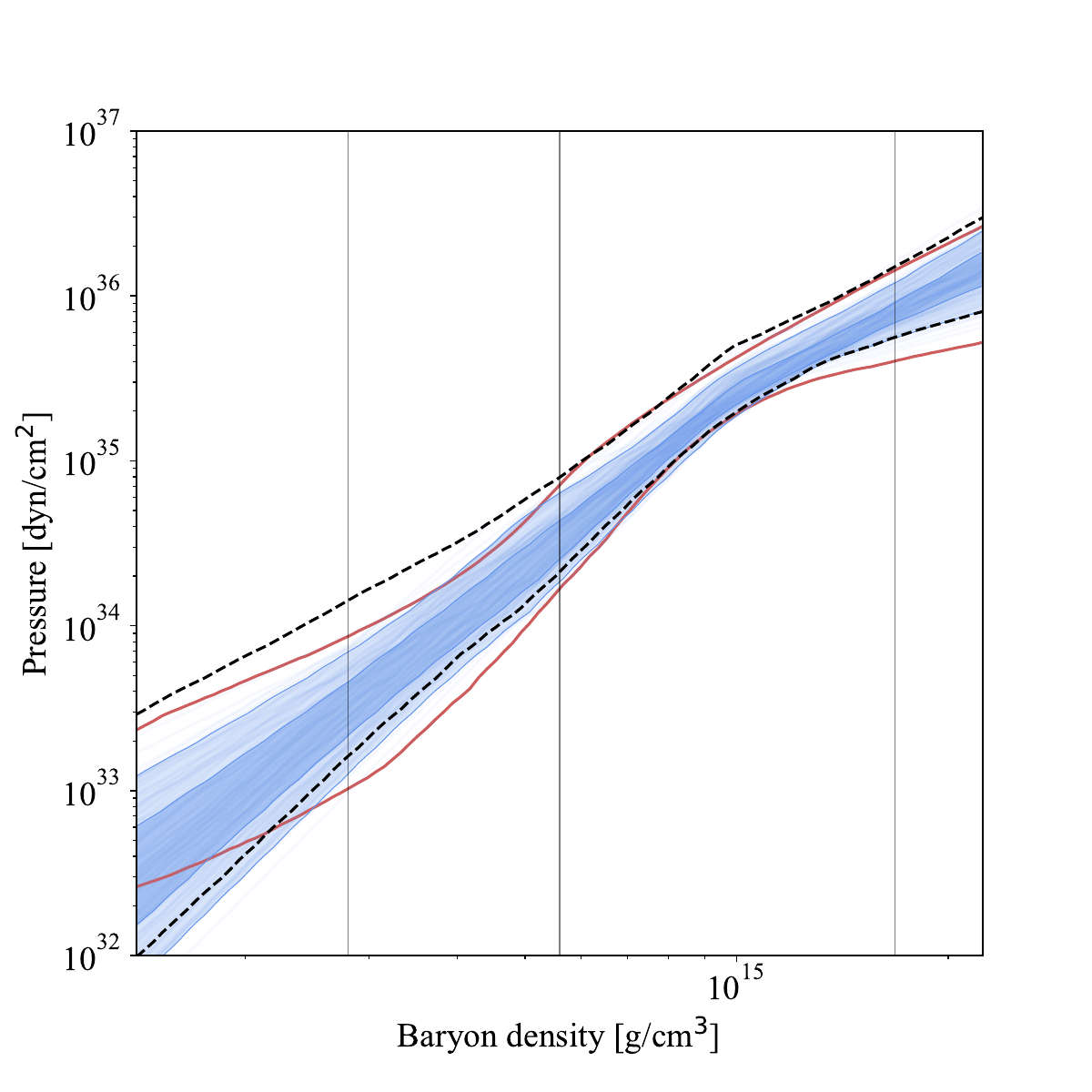}
    \caption{Constraints on the neutron star mass-radius relation (left) and equation of state (right) from our NSBH multimessenger analysis (blue) compared to those inferred using GW observations of BNS mergers and radio and X-ray pulsar observations obtained in \protect\cite{Legred:2021hdx} (red, 90\% credible interval). The shaded regions show the 50\% and 90\% credible intervals, while the dashed black lines enclose the 90\% credible region spanned by the prior on the $\boldsymbol{\Lambda}_{\mathrm{EoS}}$ parameters. The faint blue lines show individual mass-radius relation or EoS posterior draws from our NSBH analysis. The grey vertical lines in the right panel indicate once, twice and six times the nuclear saturation density of $2.8 \times 10^{14}$ g/cm$^3$.
    }
    \label{fig:mass-radius_joint}
\end{figure*}

In Fig.~\ref{fig:mass-radius_joint}, we show the constraints obtained on the neutron star mass-radius relation and EoS from the multimessenger NSBH analysis (blue) and the constrains from GW observations of BNS mergers and radio and X-ray pulsar observations from \cite{Legred:2021hdx} (red). 
Our result in blue represents a first step towards an EoS constraint from gravitational waves that self-consistently accounts for neutron star spin. 
Compared to the mass-radius relations allowed by the prior on $\boldsymbol{\Lambda}_{\mathrm{EoS}}$ (dashed black), our posterior rules out the stiffest EoSs yielding the largest radii for a given mass. Stiffer EoSs support more significant tidal disruption of the neutron star, which leads to enhanced remnant mass ejection. Thus, enforcing that there should be no remnant mass left over after the merger rules out this part of the EoS parameter space. Moreover, as can be seen in the EoS posterior in the right panel of Fig.~\ref{fig:mass-radius_joint}, the constraining power of this hierarchical multimessenger nondetection method is relatively consistent across all plausible neutron star densities, rather than being more concentrated at a specific density scale as is typical for the constraint from tidal measurements of an individual BNS merger. We find that the upper limit on the radius from the NSBH analysis is comparable to the \cite{Legred:2021hdx} analysis, while their lower limit on the radius is more constraining than ours.

\begin{figure}
	\includegraphics[width=\columnwidth]{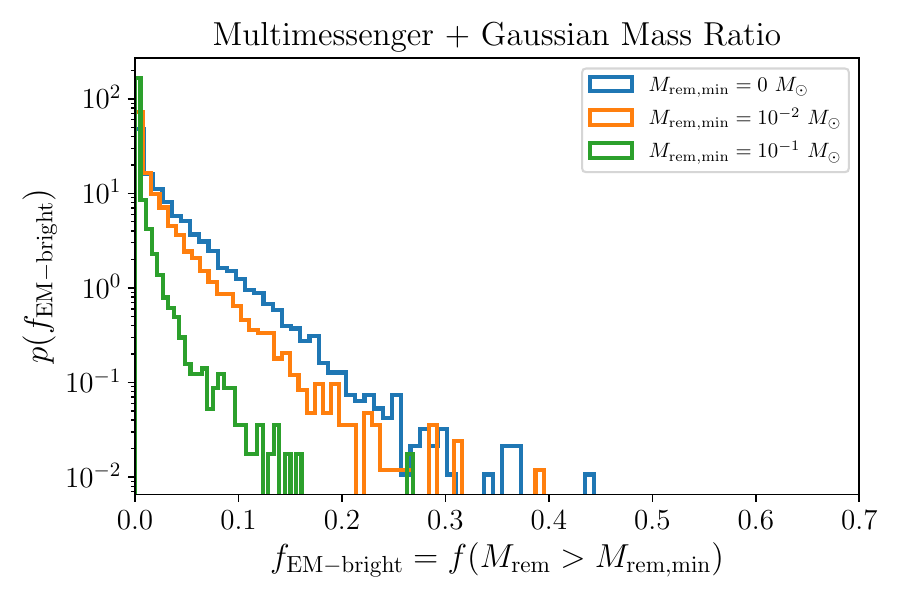}
    \caption{Posterior on the fraction of GW-detectable NSBH systems that will be electromagnetically bright with remnant mass $\hat{M}_{\mathrm{rem}} \geq M_{\mathrm{rem, min}}$, with the different colors indicating different values of $M_{\mathrm{rem, min}}$. The posterior is obtained by marginalizing over the uncertainty in both the binary population and equation of state hyperparameters under the NSBH-only multimessenger analysis that assumes none of the four detected NSBHs produced ejecta.}
    \label{fig:em_bright_joint}
\end{figure}

Finally, in Fig.~\ref{fig:em_bright_joint}, we show the posterior on the EM-bright fraction obtained using our NSBH multimessenger analysis. Rather than marginalizing over the EoS uncertainty using the \cite{legred_isaac_2022_6502467} posteriors, we use the posterior on the $\boldsymbol{\Lambda}_{\mathrm{EoS}}$ piecewise polytrope parameters obtained using the method described in Section~\ref{sec:joint_methods}. The posterior on the EM-bright fraction is similar to the one shown in the top panel of Fig.~\ref{fig:em_bright} but is more narrowly peaked at $f_{\text{EM-bright}}=0$. We find $f(\hat{M}_{\mathrm{rem}} \geq 0) \leq 0.075$ at 90\% credibility. Because our EoS constraints are comparable to those of \cite{Legred:2021hdx}, the source of this difference must be the improved measurement of the mass ratio distribution under the multimessenger model, which favors more extreme mass ratios and hence less remnant mass following the merger. In this case we find that the samples corresponding to the largest EM-bright fractions are driven by the hyperparameter posteriors that favor large black hole spins and stiff EoSs, consistent with the prediction that the NSs in these systems are more easily disrupted.

\section{Discussion}
\label{sec:discussion}
In this work, we have analyzed the population properties and multimessenger prospects of four neutron star-black hole merger events detected in gravitational waves with a false alarm rate less than 1 per year~\citep{LIGOScientific:2021djp}, GW190426\_152155, GW190917\_114630, GW200105\_162426, and GW200115\_042309. We exclude GW190814 from our analysis as the classification of the secondary object as a neutron star is unlikely. We first measured the population distributions of the black hole mass and spin and binary mass ratio using gravitational-wave data alone, marginalizing over the uncertainty in these distributions along with realistic uncertainty in the neutron star EoS to obtain a posterior on the fraction of NSBH sources detectable in gravitational waves that will be electromagnetically bright. We then developed a new Bayesian multimessenger analysis method that folds in the nondetection of any electromagnetic counterpart for these four sources to obtain constraints on the mass and spin distributions, the neutron star EoS, and the EM-bright fraction.

We find that the maximum black hole mass in NSBH systems is much lower than the maximum mass for binary black hole systems, with a distance of 24 standard deviations of the NSBH $m_{\mathrm{BH}, \max}$ measurement between them. Our measurement of the maximum mass is lower than that obtained in \cite{Zhu:2021jbw} due to their inclusion of an additional event with a more massive primary that was detected with lower FAR. Our measurement of the minimum black hole mass, on the other hand, is consistent with \cite{Zhu:2021jbw}, \citep{Ye:2022qoe}, and the minimum mass inferred from the BBH population, which all peak around $\sim 5.5~M_{\odot}$. When taken together with the inference on the maximum neutron star mass, we measure the width of the lower mass gap between the most massive neutron stars and the least massive black holes to be $3.52^{+1.15}_{-2.56}~M_{\odot}$, with no mass gap (width $\leq 0~M_{\odot}$) disfavored at 98.6\% credibility under the Gaussian pairing function, consistent with the results of \cite{Ye:2022qoe}. This inference is inherently contingent on the exclusion of GW190814 from our analysis. Our inferred neutron star mass distribution is qualitatively consistent with the result of \cite{LIGOScientific:2021psn}; the differences in minimum and maximum neutron star mass are largely attributable to different prior choices between our analyses.

Our inference on the mass ratio distribution is dominated by the prior ranges imposed on the component masses, which result in a peak at $q\sim 0.1$. We find no statistical preference between a Gaussian or power-law pairing function, as may be expected given our limited sample size of four detections. The distribution obtained under the power-law pairing function is more informative, however, and falls off more gradually from the peak. When we fold in the nondetection of any electromagnetic counterparts for the four NSBH systems in our population, the multimessenger analysis rules out a slightly larger region of the mass ratio hyperparameter space corresponding to large values of the mean and small values of the standard deviation for the Gaussian pairing function.

Unlike \cite{Ye:2022qoe} which do not explicitly fit the black hole spin distribution and \cite{Zhu:2021jbw} which fit only the effective aligned and precessing spin distributions, we fit the black hole spin magnitude distribution directly, assuming that the spin orientations are isotropically distributed. We find that the spin magnitude distribution is strongly peaked at $\chi_{\mathrm{BH}}=0$, with a maximum spin smaller than the BBH population, although some of the differences between the NSBH and BBH spin magnitude distribution can be attributed to different prior choices (see Appendix~\ref{ap:bh_spin}).

We have verified that our results are robust when using just the two events with the lowest FAR in the population, GW200105\_162426 and GW200115\_042309, as indicated in Table~\ref{tab:events}. This is expected due to the fact that the two lower-significance events have very similar masses to the two higher-significance events, as shown in Fig.~\ref{fig:1d_hists}. As such, the population parameters we infer with just two events are similar to those inferred with all four events, just less well-constrained (see Fig.~\ref{fig:low_far}). However, some of our results naturally depend on how we have defined the selected population and imposed prior knowledge of source classification. This is a feature of hierarchical inference when all analyzed candidates are assumed to be real astrophysical events belonging to a single population. For example, when we include GW190425---most likely a binary neutron star merger---in our population along with the four events previously analyzed, we find no evidence for a mass gap, with $m_{\mathrm{BH, min}} \leq 2.40~M_{\odot}$ at 90\% credibility. Correspondingly, the posterior on the EM-bright fraction shifts to higher values, with $f(\hat{M}_{\mathrm{rem}} > 0~M_{\odot}) = 0.16^{+0.13}_{-0.16}$. More details on the analysis including GW190425 are included in Appendix~\ref{ap:gw190425}. An alternative approach that we leave to future work would be to simultaneously infer the properties and classification of individual events into multiple sub-populations~\citep{Farah:2021qom}.

The differences between the mass and spin distributions of the black holes in NSBHs and BBHs may indicate that the two populations draw from different stellar progenitors and potentially form via different channels. While it is likely that some fraction of BBHs form dynamically~\citep[e.g.][]{ Romero-Shaw:2020owr, Zevin:2020gbd, Bouffanais:2021wcr}, potentially via hierarchical mergers~\citep{Kimball:2020qyd, Mould:2022ccw} in order to explain the support for masses in the ``upper mass gap'' predicted by isolated binary evolution models~\citep{Heger:2001cd, Belczynski:2016jno} along with population-level evidence for spin precession~\citep{LIGOScientific:2020kqk, LIGOScientific:2021psn}, the lower masses and smaller spin magnitudes observed in the NSBH population do not as strongly suggest a dynamical origin~\citep{Ye:2019xvf}. Concrete statements about the formation channels of NSBH systems are difficult to make with so few observations, however. A meaningful constraint on the distribution of black hole spin tilts would be particularly useful for this purpose; the posteriors for several of the candidate events we consider individually support tilts anti-aligned to the orbital angular momentum~\citep{Gompertz:2021xub}, although \citet{Mandel:2021ewy} find that this is driven by the prior choice. They conclude that a more astrophysically-motivated prior~\citep{Broekgaarden:2021hlu,Chattopadhyay:2022cnp} yields posteriors consistent with small black hole spin, as we find in this work. We leave a full hierarchical analysis of the black hole tilt distribution for NSBHs to future work when more detections can be included in the inference.

In addition to constraining the distributions of the binary mass and spin parameters, we present a data-driven estimate of the fraction of NSBH sources detectable in gravitational waves that may also have an electromagnetic counterpart. When considering the gravitational-wave NSBH data alone in conjunction with recent constraints on the neutron star EoS from gravitational-wave observations of BNS mergers and radio and X-ray pulsar observations, we find that at most 14\% of detectable sources will be left with any remnant mass outside the black hole ISCO radius that can potentially power an electromagnetic counterpart. Consistent with the small fraction of sources that have the potential to be EM-bright, we find that the maximum contribution of the astrophysical population of NSBH mergers to heavy element production in the universe is $\leq 1.03~M_{\odot}/\mathrm{Gpc^{-3}yr^{-1}}$. When we factor in the nondetection of any electromagnetic counterparts to the four NSBH mergers in our population, the EM-bright fraction drops to at most 7.5\%. Unlike previous estimates of the EM-bright fraction based on population synthesis simulations~\citep{Drozda:2020qab, Fragione:2021cvv, Broekgaarden:2021iew}, our result accounts for the measured black hole spin and mass ratio distributions, which favor small spins and extreme mass ratios and hence a lower neutron star disruption probability. We also account for neutron star spin in modeling the EoS that goes into the calculation of the remnant mass. This data-driven approach leads to a more pessimistic outlook on the likelihood of observing electromagnetic counterparts to NSBH mergers detected in gravitational waves.

Finally, our multimessenger analysis that includes the nondetection of any NSBH electromagnetic counterparts allows us to place independent constraints on the neutron star EoS. Our results represent a conservative upper limit on the constraining power of nondetection in such a multimessenger NSBH analysis, as we ignore highly uncertain electromagnetic selection effects and instead make the maximally optimistic assumption (in the case of nondetection) that no counterpart was observed because there was exactly no remnant mass left after the merger. Even in this most optimistic scenario that assumes perfect surveys with no selection effects, our EoS constraints are comparable to those obtained using existing gravitational-wave measurements of tides in BNS and radio and X-ray pulsar observations. As such, we conclude that multimessenger analyses of NSBH mergers are not a promising method for measuring the neutron star EoS. More realistic constraints for NSBH would require considerably more complicated modeling of EM selection effects and will be even less constraining than the ones we present here. Since the results obtained with the current set of NSBH detections are less constraining than the joint BNS and pulsar measurements, we expect the disparity in the constraining power of the two methods to persist or even grow as the relative number of NSBH and BNS observations remains constant due to detection rate scaling arguments. %
Even if an EM counterpart were detected for a NSBH merger, EM selection effects must still be accounted for to obtain an accurate joint multimessenger constraint on the EoS from multiple such detections. 
Taken together, our results suggest that detections of electromagnetic counterparts to NSBH mergers are likely to be rare and that the lack of detections is relatively uninformative about the EoS compared to other means of probing neutron star matter.

\section*{Acknowledgements}
The authors thank Floor Broekgaarden, Carl-Johan Haster, Elias Most, Colm Talbot, Francois Foucart, Jorge Rueda, and Hsin-Yu Chen for helpful discussions pertaining to the manuscript. We also thank the anonymous Referee for suggestions that improved our analysis.
S.B. and S.V.\ acknowledge support of the National Science Foundation and the LIGO Laboratory.
LIGO was constructed by the California Institute of Technology and
Massachusetts Institute of Technology with funding from the National
Science Foundation and operates under cooperative agreement PHY-0757058.
S.B. is also supported by the NSF Graduate Research Fellowship under Grant No. DGE-1122374.
S.V. is also supported by NSF PHY-2045740.
P.L. is supported by the Natural Sciences \& Engineering Research Council of Canada (NSERC).
The authors are grateful for computational resources provided by the LIGO Lab and supported by NSF Grants PHY-0757058 and PHY-0823459.
This paper carries LIGO document number LIGO-P2200200.

\section*{Data Availability}

We use the publicly-available individual-event posterior samples released by the LVK~\citep{gwtc2_data_release_samples, gwtc2.1_data_release, gwtc3_data_release} as input for our NSBH hierarchical inference, which is performed using the following programs:
 \textsc{Bilby}~\citep{Ashton:2018jfp, Romero-Shaw:2020owr}, \textsc{dynesty}~\citep{Speagle:2019ivv}, \textsc{nestle}~\citep{nestle}, \textsc{GWPopulation}~\citep{Talbot:2019okv}, and
 \textsc{eosinference}~\citep{eosinference}.
Our hierarchical inference results including hyperparameter posterior samples are publicly available on \href{https://doi.org/10.5281/zenodo.6795223
}{Zenodo}.

\bibliographystyle{mnras}
\bibliography{nsbh} %

\begin{thebibliography}{}
\makeatletter
\relax
\def\mn@urlcharsother{\let\do\@makeother \do\$\do\&\do\#\do\^\do\_\do\%\do\~}
\def\mn@doi{\begingroup\mn@urlcharsother \@ifnextchar [ {\mn@doi@}
  {\mn@doi@[]}}
\def\mn@doi@[#1]#2{\def\@tempa{#1}\ifx\@tempa\@empty \href
  {http://dx.doi.org/#2} {doi:#2}\else \href {http://dx.doi.org/#2} {#1}\fi
  \endgroup}
\def\mn@eprint#1#2{\mn@eprint@#1:#2::\@nil}
\def\mn@eprint@arXiv#1{\href {http://arxiv.org/abs/#1} {{\tt arXiv:#1}}}
\def\mn@eprint@dblp#1{\href {http://dblp.uni-trier.de/rec/bibtex/#1.xml}
  {dblp:#1}}
\def\mn@eprint@#1:#2:#3:#4\@nil{\def\@tempa {#1}\def\@tempb {#2}\def\@tempc
  {#3}\ifx \@tempc \@empty \let \@tempc \@tempb \let \@tempb \@tempa \fi \ifx
  \@tempb \@empty \def\@tempb {arXiv}\fi \@ifundefined
  {mn@eprint@\@tempb}{\@tempb:\@tempc}{\expandafter \expandafter \csname
  mn@eprint@\@tempb\endcsname \expandafter{\@tempc}}}

\bibitem[\protect\citeauthoryear{Aasi et~al.}{Aasi
  et~al.}{2015}]{LIGOScientific:2014pky}
Aasi J.,  et~al., 2015, \mn@doi [Class. Quant. Grav.]
  {10.1088/0264-9381/32/7/074001}, 32, 074001

\bibitem[\protect\citeauthoryear{Abbott et~al.}{Abbott
  et~al.}{2017}]{LIGOScientific:2017vwq}
Abbott B.~P.,  et~al., 2017, \mn@doi [Phys. Rev. Lett.]
  {10.1103/PhysRevLett.119.161101}, 119, 161101

\bibitem[\protect\citeauthoryear{Abbott et~al.}{Abbott
  et~al.}{2020a}]{gwtc2_data_release_samples}
Abbott B.,  et~al., 2020a, {GWTC-2 Data Release: Parameter Estimation Samples
  and Skymaps}, \url {{https://dcc.ligo.org/LIGO-P2000223/public/}}

\bibitem[\protect\citeauthoryear{Abbott et~al.}{Abbott
  et~al.}{2020b}]{LIGOScientific:2020aai}
Abbott B.~P.,  et~al., 2020b, \mn@doi [Astrophys. J. Lett.]
  {10.3847/2041-8213/ab75f5}, 892, L3

\bibitem[\protect\citeauthoryear{Abbott et~al.}{Abbott
  et~al.}{2020c}]{LIGOScientific:2020zkf}
Abbott R.,  et~al., 2020c, \mn@doi [Astrophys. J. Lett.]
  {10.3847/2041-8213/ab960f}, 896, L44

\bibitem[\protect\citeauthoryear{Abbott et~al.}{Abbott
  et~al.}{2021a}]{ligo_scientific_collaboration_and_virgo_2021_5636816}
Abbott R.,  et~al., 2021a, {GWTC-3: Compact Binary Coalescences Observed by
  LIGO and Virgo During the Second Part of the Third Observing Run — O1+O2+O3
  Search Sensitivity Estimates}, \mn@doi{10.5281/zenodo.5636816}, \url
  {https://doi.org/10.5281/zenodo.5636816}

\bibitem[\protect\citeauthoryear{Abbott et~al.}{Abbott
  et~al.}{2021b}]{LIGOScientific:2021usb}
Abbott R.,  et~al., 2021b, arXiv e-prints, \href
  {https://ui.adsabs.harvard.edu/abs/2021arXiv210801045T} {p. arXiv:2108.01045}

\bibitem[\protect\citeauthoryear{Abbott et~al.}{Abbott
  et~al.}{2021c}]{LIGOScientific:2021djp}
Abbott R.,  et~al., 2021c, arXiv e-prints, \href
  {https://ui.adsabs.harvard.edu/abs/2021arXiv211103606T} {p. arXiv:2111.03606}

\bibitem[\protect\citeauthoryear{Abbott et~al.}{Abbott
  et~al.}{2021d}]{LIGOScientific:2021psn}
Abbott R.,  et~al., 2021d, arXiv e-prints, \href
  {https://ui.adsabs.harvard.edu/abs/2021arXiv211103634T} {p. arXiv:2111.03634}

\bibitem[\protect\citeauthoryear{Abbott et~al.}{Abbott
  et~al.}{2021e}]{LIGOScientific:2020kqk}
Abbott R.,  et~al., 2021e, \mn@doi [Astrophys. J. Lett.]
  {10.3847/2041-8213/abe949}, 913, L7

\bibitem[\protect\citeauthoryear{Abbott et~al.}{Abbott
  et~al.}{2021f}]{LIGOScientific:2021qlt}
Abbott R.,  et~al., 2021f, \mn@doi [Astrophys. J. Lett.]
  {10.3847/2041-8213/ac082e}, 915, L5

\bibitem[\protect\citeauthoryear{Acernese et~al.}{Acernese
  et~al.}{2015}]{VIRGO:2014yos}
Acernese F.,  et~al., 2015, \mn@doi [Class. Quant. Grav.]
  {10.1088/0264-9381/32/2/024001}, 32, 024001

\bibitem[\protect\citeauthoryear{Antoniadis et~al.}{Antoniadis
  et~al.}{2013}]{Antoniadis:2013pzd}
Antoniadis J.,  et~al., 2013, \mn@doi [Science] {10.1126/science.1233232}, 340,
  6131

\bibitem[\protect\citeauthoryear{Ascenzi, De~Lillo, Haster, Ohme  \&
  Pannarale}{Ascenzi et~al.}{2019}]{Ascenzi:2018mwp}
Ascenzi S.,  De~Lillo N.,  Haster C.-J.,  Ohme F.,   Pannarale F.,  2019,
  \mn@doi [Astrophys. J.] {10.3847/1538-4357/ab1b15}, 877, 94

\bibitem[\protect\citeauthoryear{Ashton et~al.}{Ashton
  et~al.}{2019}]{Ashton:2018jfp}
Ashton G.,  et~al., 2019, \mn@doi [Astrophys. J. Suppl.]
  {10.3847/1538-4365/ab06fc}, 241, 27

\bibitem[\protect\citeauthoryear{Barbary et~al.}{Barbary et~al.}{2021}]{nestle}
Barbary K.,  et~al., 2021, Nestle, \url{https://github.com/kbarbary/nestle}

\bibitem[\protect\citeauthoryear{Barbieri, Salafia, Perego, Colpi  \&
  Ghirlanda}{Barbieri et~al.}{2019}]{Barbieri:2019sjc}
Barbieri C.,  Salafia O.~S.,  Perego A.,  Colpi M.,   Ghirlanda G.,  2019,
  \mn@doi [Astron. Astrophys.] {10.1051/0004-6361/201935443}, 625, A152

\bibitem[\protect\citeauthoryear{{Bavera}, {Zevin}  \& {Fragos}}{{Bavera}
  et~al.}{2021}]{Bavera:2021evk}
{Bavera} S.~S.,  {Zevin} M.,   {Fragos} T.,  2021, \mn@doi [Research Notes of
  the American Astronomical Society] {10.3847/2515-5172/ac053c}, \href
  {https://ui.adsabs.harvard.edu/abs/2021RNAAS...5..127B} {5, 127}

\bibitem[\protect\citeauthoryear{Belczynski et~al.}{Belczynski
  et~al.}{2016}]{Belczynski:2016jno}
Belczynski K.,  et~al., 2016, \mn@doi [Astron. Astrophys.]
  {10.1051/0004-6361/201628980}, 594, A97

\bibitem[\protect\citeauthoryear{Biscoveanu, Talbot  \& Vitale}{Biscoveanu
  et~al.}{2022}]{Biscoveanu:2021eht}
Biscoveanu S.,  Talbot C.,   Vitale S.,  2022, \mn@doi [Mon. Not. Roy. Astron.
  Soc.] {10.1093/mnras/stac347}, 511, 4350

\bibitem[\protect\citeauthoryear{Bouffanais, Mapelli, Santoliquido, Giacobbo,
  Di~Carlo, Rastello, Artale  \& Iorio}{Bouffanais
  et~al.}{2021}]{Bouffanais:2021wcr}
Bouffanais Y.,  Mapelli M.,  Santoliquido F.,  Giacobbo N.,  Di~Carlo U.~N.,
  Rastello S.,  Artale M.~C.,   Iorio G.,  2021, \mn@doi [Mon. Not. Roy.
  Astron. Soc.] {10.1093/mnras/stab2438}, 507, 5224

\bibitem[\protect\citeauthoryear{Breu \& Rezzolla}{Breu \&
  Rezzolla}{2016}]{Breu:2016ufb}
Breu C.,  Rezzolla L.,  2016, \mn@doi [Mon. Not. Roy. Astron. Soc.]
  {10.1093/mnras/stw575}, 459, 646

\bibitem[\protect\citeauthoryear{Broekgaarden \& Berger}{Broekgaarden \&
  Berger}{2021}]{Broekgaarden:2021hlu}
Broekgaarden F.~S.,  Berger E.,  2021, \mn@doi [Astrophys. J. Lett.]
  {10.3847/2041-8213/ac2832}, 920, L13

\bibitem[\protect\citeauthoryear{Broekgaarden et~al.,}{Broekgaarden
  et~al.}{2021}]{Broekgaarden:2021iew}
Broekgaarden F.~S.,  et~al., 2021, \mn@doi [Mon. Not. Roy. Astron. Soc.]
  {10.1093/mnras/stab2716}, 508, 5028

\bibitem[\protect\citeauthoryear{Chase et~al.,}{Chase
  et~al.}{2022}]{Chase:2021ood}
Chase E.~A.,  et~al., 2022, \mn@doi [Astrophys. J.] {10.3847/1538-4357/ac3d25},
  927, 163

\bibitem[\protect\citeauthoryear{Chattopadhyay, Stevenson, Hurley, Bailes  \&
  Broekgaarden}{Chattopadhyay et~al.}{2021}]{Chattopadhyay:2020lff}
Chattopadhyay D.,  Stevenson S.,  Hurley J.~R.,  Bailes M.,   Broekgaarden F.,
  2021, \mn@doi [Mon. Not. Roy. Astron. Soc.] {10.1093/mnras/stab973}, 504,
  3682

\bibitem[\protect\citeauthoryear{Chattopadhyay, Stevenson, Broekgaarden,
  Antonini  \& Belczynski}{Chattopadhyay et~al.}{2022}]{Chattopadhyay:2022cnp}
Chattopadhyay D.,  Stevenson S.,  Broekgaarden F.,  Antonini F.,   Belczynski
  K.,  2022, \mn@doi [Mon. Not. Roy. Astron. Soc.] {10.1093/mnras/stac1283},
  531, 5780

\bibitem[\protect\citeauthoryear{Chen, Vitale  \& Foucart}{Chen
  et~al.}{2021}]{Chen:2021fro}
Chen H.-Y.,  Vitale S.,   Foucart F.,  2021, \mn@doi [Astrophys. J. Lett.]
  {10.3847/2041-8213/ac26c6}, 920, L3

\bibitem[\protect\citeauthoryear{Cipolletta, Cherubini, Filippi, Rueda  \&
  Ruffini}{Cipolletta et~al.}{2015}]{Cipolletta:2015nga}
Cipolletta F.,  Cherubini C.,  Filippi S.,  Rueda J.~A.,   Ruffini R.,  2015,
  \mn@doi [Phys. Rev. D] {10.1103/PhysRevD.92.023007}, 92, 023007

\bibitem[\protect\citeauthoryear{Collaboration \& Collaboration}{Collaboration
  \& Collaboration}{2021}]{gwtc2.1_data_release}
Collaboration L.~S.,  Collaboration V.,  2021, {GWTC-2.1: Deep Extended Catalog
  of Compact Binary Coalescences Observed by LIGO and Virgo During the First
  Half of the Third Observing Run - Parameter Estimation Data Release},
  \mn@doi{10.5281/zenodo.5117703}, \url
  {https://doi.org/10.5281/zenodo.5117703}

\bibitem[\protect\citeauthoryear{Collaboration, Collaboration  \&
  Collaboration}{Collaboration et~al.}{2021}]{gwtc3_data_release}
Collaboration L.~S.,  Collaboration V.,   Collaboration K.,  2021, {GWTC-3:
  Compact Binary Coalescences Observed by LIGO and Virgo During the Second Part
  of the Third Observing Run — Parameter estimation data release},
  \mn@doi{10.5281/zenodo.5546663}, \url
  {https://doi.org/10.5281/zenodo.5546663}

\bibitem[\protect\citeauthoryear{C\^ot\'e et~al.}{C\^ot\'e
  et~al.}{2018}]{Cote:2017evr}
C\^ot\'e B.,  et~al., 2018, \mn@doi [Astrophys. J.] {10.3847/1538-4357/aaad67},
  855, 99

\bibitem[\protect\citeauthoryear{Coughlin \& Dietrich}{Coughlin \&
  Dietrich}{2019}]{Coughlin:2019kqf}
Coughlin M.~W.,  Dietrich T.,  2019, \mn@doi [Phys. Rev. D]
  {10.1103/PhysRevD.100.043011}, 100, 043011

\bibitem[\protect\citeauthoryear{Coughlin, Dietrich, Margalit  \&
  Metzger}{Coughlin et~al.}{2019}]{Coughlin:2018fis}
Coughlin M.~W.,  Dietrich T.,  Margalit B.,   Metzger B.~D.,  2019, \mn@doi
  [Mon. Not. Roy. Astron. Soc.] {10.1093/mnrasl/slz133}, 489, L91

\bibitem[\protect\citeauthoryear{Cromartie et~al.}{Cromartie
  et~al.}{2019}]{NANOGrav:2019jur}
Cromartie H.~T.,  et~al., 2019, \mn@doi [Nature Astron.]
  {10.1038/s41550-019-0880-2}, 4, 72

\bibitem[\protect\citeauthoryear{{Drozda}, {Belczynski}, {O'Shaughnessy},
  {Bulik}  \& {Fryer}}{{Drozda} et~al.}{2020}]{Drozda:2020qab}
{Drozda} P.,  {Belczynski} K.,  {O'Shaughnessy} R.,  {Bulik} T.,   {Fryer}
  C.~L.,  2020, arXiv e-prints, \href
  {https://ui.adsabs.harvard.edu/abs/2020arXiv200906655D} {p. arXiv:2009.06655}

\bibitem[\protect\citeauthoryear{{Essick} \& {Landry}}{{Essick} \&
  {Landry}}{2020}]{EssickLandry2020}
{Essick} R.,  {Landry} P.,  2020, \mn@doi [ApJ] {10.3847/1538-4357/abbd3b},
  \href {https://ui.adsabs.harvard.edu/abs/2020ApJ...904...80E} {904, 80}

\bibitem[\protect\citeauthoryear{{Farah}, {Fishbach}, {Essick}, {Holz}  \&
  {Galaudage}}{{Farah} et~al.}{2021}]{Farah:2021qom}
{Farah} A.~M.,  {Fishbach} M.,  {Essick} R.,  {Holz} D.~E.,   {Galaudage} S.,
  2021, arXiv e-prints, \href
  {https://ui.adsabs.harvard.edu/abs/2021arXiv211103498F} {p. arXiv:2111.03498}

\bibitem[\protect\citeauthoryear{Farr}{Farr}{2019}]{Farr_2019}
Farr W.~M.,  2019, \mn@doi [Research Notes of the {AAS}]
  {10.3847/2515-5172/ab1d5f}, 3, 66

\bibitem[\protect\citeauthoryear{Feeney, Peiris, Nissanke  \& Mortlock}{Feeney
  et~al.}{2021}]{Feeney:2020kxk}
Feeney S.~M.,  Peiris H.~V.,  Nissanke S.~M.,   Mortlock D.~J.,  2021, \mn@doi
  [Phys. Rev. Lett.] {10.1103/PhysRevLett.126.171102}, 126, 171102

\bibitem[\protect\citeauthoryear{Fern\'andez, Foucart, Kasen, Lippuner, Desai
  \& Roberts}{Fern\'andez et~al.}{2017}]{Fernandez:2016sbf}
Fern\'andez R.,  Foucart F.,  Kasen D.,  Lippuner J.,  Desai D.,   Roberts
  L.~F.,  2017, \mn@doi [Class. Quant. Grav.] {10.1088/1361-6382/aa7a77}, 34,
  154001

\bibitem[\protect\citeauthoryear{Fishbach \& Holz}{Fishbach \&
  Holz}{2017}]{Fishbach:2017zga}
Fishbach M.,  Holz D.~E.,  2017, \mn@doi [Astrophys. J. Lett.]
  {10.3847/2041-8213/aa9bf6}, 851, L25

\bibitem[\protect\citeauthoryear{Fishbach \& Holz}{Fishbach \&
  Holz}{2020}]{Fishbach:2019bbm}
Fishbach M.,  Holz D.~E.,  2020, \mn@doi [Astrophys. J. Lett.]
  {10.3847/2041-8213/ab7247}, 891, L27

\bibitem[\protect\citeauthoryear{Fonseca et~al.}{Fonseca
  et~al.}{2021}]{Fonseca:2021wxt}
Fonseca E.,  et~al., 2021, \mn@doi [Astrophys. J. Lett.]
  {10.3847/2041-8213/ac03b8}, 915, L12

\bibitem[\protect\citeauthoryear{Foucart}{Foucart}{2012}]{Foucart:2012nc}
Foucart F.,  2012, \mn@doi [Phys. Rev. D] {10.1103/PhysRevD.86.124007}, 86,
  124007

\bibitem[\protect\citeauthoryear{Foucart, Hinderer  \& Nissanke}{Foucart
  et~al.}{2018}]{Foucart:2018rjc}
Foucart F.,  Hinderer T.,   Nissanke S.,  2018, \mn@doi [Phys. Rev. D]
  {10.1103/PhysRevD.98.081501}, 98, 081501

\bibitem[\protect\citeauthoryear{Fragione}{Fragione}{2021}]{Fragione:2021cvv}
Fragione G.,  2021, \mn@doi [Astrophys. J. Lett.] {10.3847/2041-8213/ac3bcd},
  923, L2

\bibitem[\protect\citeauthoryear{Fuller \& Ma}{Fuller \&
  Ma}{2019}]{Fuller:2019sxi}
Fuller J.,  Ma L.,  2019, \mn@doi [Astrophys. J. Lett.]
  {10.3847/2041-8213/ab339b}, 881, L1

\bibitem[\protect\citeauthoryear{{Fuller}, {Piro}  \& {Jermyn}}{{Fuller}
  et~al.}{2019}]{2019MNRAS.485.3661F}
{Fuller} J.,  {Piro} A.~L.,   {Jermyn} A.~S.,  2019, \mn@doi [\mnras]
  {10.1093/mnras/stz514}, \href
  {https://ui.adsabs.harvard.edu/abs/2019MNRAS.485.3661F} {485, 3661}

\bibitem[\protect\citeauthoryear{Gompertz, Nicholl, Schmidt, Pratten  \&
  Vecchio}{Gompertz et~al.}{2022}]{Gompertz:2021xub}
Gompertz B.~P.,  Nicholl M.,  Schmidt P.,  Pratten G.,   Vecchio A.,  2022,
  \mn@doi [Mon. Not. Roy. Astron. Soc.] {10.1093/mnras/stac029}, 511, 1454

\bibitem[\protect\citeauthoryear{Heger \& Woosley}{Heger \&
  Woosley}{2002}]{Heger:2001cd}
Heger A.,  Woosley S.~E.,  2002, \mn@doi [Astrophys. J.] {10.1086/338487}, 567,
  532

\bibitem[\protect\citeauthoryear{Hernandez~Vivanco, Smith, Thrane, Lasky,
  Talbot  \& Raymond}{Hernandez~Vivanco
  et~al.}{2019}]{HernandezVivanco:2019vvk}
Hernandez~Vivanco F.,  Smith R.,  Thrane E.,  Lasky P.~D.,  Talbot C.,
  Raymond V.,  2019, \mn@doi [Phys. Rev. D] {10.1103/PhysRevD.100.103009}, 100,
  103009

\bibitem[\protect\citeauthoryear{Hinderer et~al.}{Hinderer
  et~al.}{2019}]{Hinderer:2018pei}
Hinderer T.,  et~al., 2019, \mn@doi [Phys. Rev. D]
  {10.1103/PhysRevD.100.063021}, 100, 06321

\bibitem[\protect\citeauthoryear{Hu, Zhu, Qin, Zhang, Liang  \& Shao}{Hu
  et~al.}{2022}]{Hu:2022ubh}
Hu R.-C.,  Zhu J.-P.,  Qin Y.,  Zhang B.,  Liang E.-W.,   Shao Y.,  2022,
  \mn@doi [Astrophys. J.] {10.3847/1538-4357/ac573f}, 928, 163

\bibitem[\protect\citeauthoryear{Huang, Haster, Vitale, Varma, Foucart  \&
  Biscoveanu}{Huang et~al.}{2021}]{Huang:2020pba}
Huang Y.,  Haster C.-J.,  Vitale S.,  Varma V.,  Foucart F.,   Biscoveanu S.,
  2021, \mn@doi [Phys. Rev. D] {10.1103/PhysRevD.103.083001}, 103, 083001

\bibitem[\protect\citeauthoryear{Janka, Eberl, Ruffert  \& Fryer}{Janka
  et~al.}{1999}]{Janka:1999qu}
Janka H.~T.,  Eberl T.,  Ruffert M.,   Fryer C.~L.,  1999, \mn@doi [Astrophys.
  J. Lett.] {10.1086/312397}, 527, L39

\bibitem[\protect\citeauthoryear{Kawaguchi, Kyutoku, Shibata  \&
  Tanaka}{Kawaguchi et~al.}{2016}]{Kawaguchi:2016ana}
Kawaguchi K.,  Kyutoku K.,  Shibata M.,   Tanaka M.,  2016, \mn@doi [Astrophys.
  J.] {10.3847/0004-637X/825/1/52}, 825, 52

\bibitem[\protect\citeauthoryear{Kimball et~al.}{Kimball
  et~al.}{2021}]{Kimball:2020qyd}
Kimball C.,  et~al., 2021, \mn@doi [Astrophys. J. Lett.]
  {10.3847/2041-8213/ac0aef}, 915, L35

\bibitem[\protect\citeauthoryear{Koliogiannis \& Moustakidis}{Koliogiannis \&
  Moustakidis}{2020}]{Koliogiannis:2019rvh}
Koliogiannis P.~S.,  Moustakidis C.~C.,  2020, \mn@doi [Phys. Rev. C]
  {10.1103/PhysRevC.101.015805}, 101, 015805

\bibitem[\protect\citeauthoryear{{Konstantinou} \& {Morsink}}{{Konstantinou} \&
  {Morsink}}{2022}]{Konstantinou:2022vkr}
{Konstantinou} A.,  {Morsink} S.~M.,  2022, arXiv e-prints, \href
  {https://ui.adsabs.harvard.edu/abs/2022arXiv220612515K} {p. arXiv:2206.12515}

\bibitem[\protect\citeauthoryear{Lackey}{Lackey}{2019}]{eosinference}
Lackey B.,  2019, eosinference,
  \url{https://github.com/benjaminlackey/eosinference}

\bibitem[\protect\citeauthoryear{Lackey \& Wade}{Lackey \&
  Wade}{2015}]{Lackey:2014fwa}
Lackey B.~D.,  Wade L.,  2015, \mn@doi [Phys. Rev. D]
  {10.1103/PhysRevD.91.043002}, 91, 043002

\bibitem[\protect\citeauthoryear{Lackey, Kyutoku, Shibata, Brady  \&
  Friedman}{Lackey et~al.}{2012}]{Lackey:2011vz}
Lackey B.~D.,  Kyutoku K.,  Shibata M.,  Brady P.~R.,   Friedman J.~L.,  2012,
  \mn@doi [Phys. Rev. D] {10.1103/PhysRevD.85.044061}, 85, 044061

\bibitem[\protect\citeauthoryear{Landry \& Read}{Landry \&
  Read}{2021}]{Landry:2021hvl}
Landry P.,  Read J.~S.,  2021, \mn@doi [Astrophys. J. Lett.]
  {10.3847/2041-8213/ac2f3e}, 921, L25

\bibitem[\protect\citeauthoryear{Legred, Chatziioannou, Essick, Han  \&
  Landry}{Legred et~al.}{2021}]{Legred:2021hdx}
Legred I.,  Chatziioannou K.,  Essick R.,  Han S.,   Landry P.,  2021, \mn@doi
  [Phys. Rev. D] {10.1103/PhysRevD.104.063003}, 104, 063003

\bibitem[\protect\citeauthoryear{Legred, Chatziioannou, Essick, Han  \&
  Landry}{Legred et~al.}{2022}]{legred_isaac_2022_6502467}
Legred I.,  Chatziioannou K.,  Essick R.,  Han S.,   Landry P.,  2022, {Impact
  of the PSR J0740+6620 radius constraint on the properties of high-density
  matter: Neutron star equation of state posterior samples},
  \mn@doi{10.5281/zenodo.6502467}, \url
  {https://doi.org/10.5281/zenodo.6502467}

\bibitem[\protect\citeauthoryear{Loredo}{Loredo}{2004}]{Loredo_2004}
Loredo T.~J.,  2004, in Bayesian Inference and Maximum Entropy Methods in
  Science and Engineering ed R. Fischer, R. Preuss and U. V. Toussaint. AIP,
  Melville, NY, p.~195, \mn@doi{10.1063/1.1835214}

\bibitem[\protect\citeauthoryear{Mandel \& Broekgaarden}{Mandel \&
  Broekgaarden}{2022}]{Mandel:2021smh}
Mandel I.,  Broekgaarden F.~S.,  2022, \mn@doi [Living Rev. Rel.]
  {10.1007/s41114-021-00034-3}, 25, 1

\bibitem[\protect\citeauthoryear{Mandel \& Smith}{Mandel \&
  Smith}{2021}]{Mandel:2021ewy}
Mandel I.,  Smith R. J.~E.,  2021, \mn@doi [Astrophys. J. Lett.]
  {10.3847/2041-8213/ac35dd}, 922, L14

\bibitem[\protect\citeauthoryear{Mandel, Farr  \& Gair}{Mandel
  et~al.}{2019}]{Mandel:2018mve}
Mandel I.,  Farr W.~M.,   Gair J.~R.,  2019, \mn@doi [Mon. Not. Roy. Astron.
  Soc.] {10.1093/mnras/stz896}, 486, 1086

\bibitem[\protect\citeauthoryear{Metzger \& Berger}{Metzger \&
  Berger}{2012}]{Metzger:2011bv}
Metzger B.~D.,  Berger E.,  2012, \mn@doi [Astrophys. J.]
  {10.1088/0004-637X/746/1/48}, 746, 48

\bibitem[\protect\citeauthoryear{{Mochkovitch}, {Hernanz}, {Isern}  \&
  {Martin}}{{Mochkovitch} et~al.}{1993}]{1993Natur.361..236M}
{Mochkovitch} R.,  {Hernanz} M.,  {Isern} J.,   {Martin} X.,  1993, \mn@doi
  [\nat] {10.1038/361236a0}, \href
  {https://ui.adsabs.harvard.edu/abs/1993Natur.361..236M} {361, 236}

\bibitem[\protect\citeauthoryear{Most, Weih  \& Rezzolla}{Most
  et~al.}{2020a}]{Most:2020kyx}
Most E.~R.,  Weih L.~R.,   Rezzolla L.,  2020a, \mn@doi [Mon. Not. Roy. Astron.
  Soc.] {10.1093/mnrasl/slaa079}, 496, L16

\bibitem[\protect\citeauthoryear{Most, Papenfort, Weih  \& Rezzolla}{Most
  et~al.}{2020b}]{Most:2020bba}
Most E.~R.,  Papenfort L.~J.,  Weih L.~R.,   Rezzolla L.,  2020b, \mn@doi [Mon.
  Not. Roy. Astron. Soc.] {10.1093/mnrasl/slaa168}, 499, L82

\bibitem[\protect\citeauthoryear{{Mould}, {Gerosa}  \& {Taylor}}{{Mould}
  et~al.}{2022}]{Mould:2022ccw}
{Mould} M.,  {Gerosa} D.,   {Taylor} S.~R.,  2022, arXiv e-prints, \href
  {https://ui.adsabs.harvard.edu/abs/2022arXiv220303651M} {p. arXiv:2203.03651}

\bibitem[\protect\citeauthoryear{Pankow, Rizzo, Rao, Berry  \& Kalogera}{Pankow
  et~al.}{2020}]{Pankow:2019oxl}
Pankow C.,  Rizzo M.,  Rao K.,  Berry C. P.~L.,   Kalogera V.,  2020, \mn@doi
  [Astrophys. J.] {10.3847/1538-4357/abb373}, 902, 71

\bibitem[\protect\citeauthoryear{Pannarale, Tonita  \& Rezzolla}{Pannarale
  et~al.}{2011}]{Pannarale:2010vs}
Pannarale F.,  Tonita A.,   Rezzolla L.,  2011, \mn@doi [Astrophys. J.]
  {10.1088/0004-637X/727/2/95}, 727, 95

\bibitem[\protect\citeauthoryear{Paschalidis, Ruiz  \& Shapiro}{Paschalidis
  et~al.}{2015}]{Paschalidis:2014qra}
Paschalidis V.,  Ruiz M.,   Shapiro S.~L.,  2015, \mn@doi [Astrophys. J. Lett.]
  {10.1088/2041-8205/806/1/L14}, 806, L14

\bibitem[\protect\citeauthoryear{Qin, Fragos, Meynet, Andrews, S\o{}rensen  \&
  Song}{Qin et~al.}{2018}]{Qin:2018vaa}
Qin Y.,  Fragos T.,  Meynet G.,  Andrews J.,  S\o{}rensen M.,   Song H.~F.,
  2018, \mn@doi [Astron. Astrophys.] {10.1051/0004-6361/201832839}, 616, A28

\bibitem[\protect\citeauthoryear{Raaijmakers et~al.}{Raaijmakers
  et~al.}{2021}]{Raaijmakers:2021slr}
Raaijmakers G.,  et~al., 2021, \mn@doi [Astrophys. J.]
  {10.3847/1538-4357/ac222d}, 922, 269

\bibitem[\protect\citeauthoryear{Read, Lackey, Owen  \& Friedman}{Read
  et~al.}{2009}]{Read:2008iy}
Read J.~S.,  Lackey B.~D.,  Owen B.~J.,   Friedman J.~L.,  2009, \mn@doi [Phys.
  Rev. D] {10.1103/PhysRevD.79.124032}, 79, 124032

\bibitem[\protect\citeauthoryear{Reynolds}{Reynolds}{2021}]{Reynolds:2020jwt}
Reynolds C.~S.,  2021, \mn@doi [Ann. Rev. Astron. Astrophys.]
  {10.1146/annurev-astro-112420-035022}, 59, 117

\bibitem[\protect\citeauthoryear{Rom\'an-Garza et~al.}{Rom\'an-Garza
  et~al.}{2021}]{Roman-Garza:2020uou}
Rom\'an-Garza J.,  et~al., 2021, \mn@doi [Astrophys. J. Lett.]
  {10.3847/2041-8213/abf42c}, 912, L23

\bibitem[\protect\citeauthoryear{Romero-Shaw et~al.}{Romero-Shaw
  et~al.}{2020}]{Romero-Shaw:2020owr}
Romero-Shaw I.~M.,  et~al., 2020, \mn@doi [Mon. Not. Roy. Astron. Soc.]
  {10.1093/mnras/staa2850}, 499, 3295

\bibitem[\protect\citeauthoryear{Ruiz, Shapiro  \& Tsokaros}{Ruiz
  et~al.}{2018}]{Ruiz:2018wah}
Ruiz M.,  Shapiro S.~L.,   Tsokaros A.,  2018, \mn@doi [Phys. Rev. D]
  {10.1103/PhysRevD.98.123017}, 98, 123017

\bibitem[\protect\citeauthoryear{Sarin, Lasky, Vivanco, Stevenson,
  Chattopadhyay, Smith  \& Thrane}{Sarin et~al.}{2022}]{Sarin:2022cmu}
Sarin N.,  Lasky P.~D.,  Vivanco F.~H.,  Stevenson S.~P.,  Chattopadhyay D.,
  Smith R.,   Thrane E.,  2022, \mn@doi [Phys. Rev. D]
  {10.1103/PhysRevD.105.083004}, 105, 083004

\bibitem[\protect\citeauthoryear{Shao, Tang, Sheng, Jiang, Wang, Jin, Fan  \&
  Wei}{Shao et~al.}{2020}]{Shao:2019ioq}
Shao D.-S.,  Tang S.-P.,  Sheng X.,  Jiang J.-L.,  Wang Y.-Z.,  Jin Z.-P.,  Fan
  Y.-Z.,   Wei D.-M.,  2020, \mn@doi [Phys. Rev. D]
  {10.1103/PhysRevD.101.063029}, 101, 063029

\bibitem[\protect\citeauthoryear{Shapiro}{Shapiro}{2017}]{Shapiro:2017cny}
Shapiro S.~L.,  2017, \mn@doi [Phys. Rev. D] {10.1103/PhysRevD.95.101303}, 95,
  101303

\bibitem[\protect\citeauthoryear{Speagle}{Speagle}{2020}]{Speagle:2019ivv}
Speagle J.~S.,  2020, \mn@doi [Mon. Not. Roy. Astron. Soc.]
  {10.1093/mnras/staa278}, 493, 3132

\bibitem[\protect\citeauthoryear{Spruit}{Spruit}{2002}]{Spruit:2001tz}
Spruit H.~C.,  2002, \mn@doi [Astron. Astrophys.] {10.1051/0004-6361:20011465},
  381, 923

\bibitem[\protect\citeauthoryear{Talbot, Smith, Thrane  \& Poole}{Talbot
  et~al.}{2019}]{Talbot:2019okv}
Talbot C.,  Smith R.,  Thrane E.,   Poole G.~B.,  2019, \mn@doi [Phys. Rev. D]
  {10.1103/PhysRevD.100.043030}, 100, 043030

\bibitem[\protect\citeauthoryear{Tanaka \& Hotokezaka}{Tanaka \&
  Hotokezaka}{2013}]{Tanaka:2013ana}
Tanaka M.,  Hotokezaka K.,  2013, \mn@doi [Astrophys. J.]
  {10.1088/0004-637X/775/2/113}, 775, 113

\bibitem[\protect\citeauthoryear{Tanaka, Hotokezaka, Kyutoku, Wanajo, Kiuchi,
  Sekiguchi  \& Shibata}{Tanaka et~al.}{2014}]{Tanaka:2013ixa}
Tanaka M.,  Hotokezaka K.,  Kyutoku K.,  Wanajo S.,  Kiuchi K.,  Sekiguchi Y.,
   Shibata M.,  2014, \mn@doi [Astrophys. J.] {10.1088/0004-637X/780/1/31},
  780, 31

\bibitem[\protect\citeauthoryear{Tang, Li, Wang, Fan  \& Wei}{Tang
  et~al.}{2021}]{Tang:2021bnp}
Tang S.-P.,  Li Y.-J.,  Wang Y.-Z.,  Fan Y.-Z.,   Wei D.-M.,  2021, \mn@doi
  [Astrophys. J.] {10.3847/1538-4357/ac22aa}, 922, 3

\bibitem[\protect\citeauthoryear{Thrane \& Talbot}{Thrane \&
  Talbot}{2019}]{Thrane:2018qnx}
Thrane E.,  Talbot C.,  2019, \mn@doi [Publ. Astron. Soc. Austral.]
  {10.1017/pasa.2019.2}, 36, e010

\bibitem[\protect\citeauthoryear{{Vitale}, {Gerosa}, {Farr}  \&
  {Taylor}}{{Vitale} et~al.}{2020}]{Vitale:2020aaz}
{Vitale} S.,  {Gerosa} D.,  {Farr} W.~M.,   {Taylor} S.~R.,  2020, arXiv
  e-prints, \href {https://ui.adsabs.harvard.edu/abs/2020arXiv200705579V} {p.
  arXiv:2007.05579}

\bibitem[\protect\citeauthoryear{Wysocki, Lange  \& O'Shaughnessy}{Wysocki
  et~al.}{2019}]{Wysocki:2018mpo}
Wysocki D.,  Lange J.,   O'Shaughnessy R.,  2019, \mn@doi [Phys. Rev. D]
  {10.1103/PhysRevD.100.043012}, 100, 043012

\bibitem[\protect\citeauthoryear{{Ye} \& {Fishbach}}{{Ye} \&
  {Fishbach}}{2022}]{Ye:2022qoe}
{Ye} C.,  {Fishbach} M.,  2022, arXiv e-prints, \href
  {https://ui.adsabs.harvard.edu/abs/2022arXiv220205164Y} {p. arXiv:2202.05164}

\bibitem[\protect\citeauthoryear{Ye, Fong, Kremer, Rodriguez, Chatterjee,
  Fragione  \& Rasio}{Ye et~al.}{2020}]{Ye:2019xvf}
Ye C.~S.,  Fong W.-f.,  Kremer K.,  Rodriguez C.~L.,  Chatterjee S.,  Fragione
  G.,   Rasio F.~A.,  2020, \mn@doi [Astrophys. J. Lett.]
  {10.3847/2041-8213/ab5dc5}, 888, L10

\bibitem[\protect\citeauthoryear{Zevin et~al.,}{Zevin
  et~al.}{2021}]{Zevin:2020gbd}
Zevin M.,  et~al., 2021, \mn@doi [Astrophys. J.] {10.3847/1538-4357/abe40e},
  910, 152

\bibitem[\protect\citeauthoryear{Zhu, Wu, Qin, Zhang, Gao  \& Cao}{Zhu
  et~al.}{2022}]{Zhu:2021jbw}
Zhu J.-P.,  Wu S.,  Qin Y.,  Zhang B.,  Gao H.,   Cao Z.,  2022, \mn@doi
  [Astrophys. J.] {10.3847/1538-4357/ac540c}, 928, 167

\makeatother
\end{thebibliography}

\appendix
\section{Derivation of the multimessenger likelihood}
\label{ap:math}
The joint likelihood of observing a particular NSBH merger event with gravitational-wave data, $d$, and remnant mass measurement $M_{\mathrm{rem}}$, is the product of the likelihoods of making each of those observations individually. We use $\boldsymbol{\theta}$ to refer to the full set of binary parameters needed to characterize the gravitational-wave emission and $\mathbf{x}$ to refer to the subset of those parameters that are also needed to characterize the remnant mass measurement, $\mathbf{x} = (q, m_{\mathrm{NS}}, \chi_{\mathrm{NS}}, \chi_{\mathrm{BH},z})$. We do not include the neutron star tidal deformability among the $\boldsymbol{\theta}$ parameters, as the waveform models we choose for $h_{k}(\boldsymbol{\theta})$ do not include this effect. 
\begin{align}
    p(d, M_{\mathrm{rem}} | \boldsymbol{\theta}, \boldsymbol{\Lambda}_{\mathrm{EoS}})
    &= p(d | \boldsymbol{\theta} )p(M_{\mathrm{rem}} | \mathbf{x},\boldsymbol{\Lambda}_{\mathrm{EoS}})\\ \nonumber
    &= p(d | \boldsymbol{\theta} )\delta(\hat{M}_{\mathrm{rem}}(\mathbf{x}, \boldsymbol{\Lambda}_{\mathrm{EoS}})).
\end{align}
The remnant mass likelihood is a delta function at $M_{\mathrm{rem}}=0~M_{\odot}$ because we explicitly make the assumption that there was precisely no remnant mass left after the merger due to the nondetection of any electromagnetic counterpart. A more realistic analysis would relax this assumption and take into account the uncertainty in the remnant mass due to various electromagnetic selection effects including telescope sensitivity and uncertainty in the brightness of the emission. We emphasize that we make this simplifying assumption in order to present the most optimistic multimessenger constraints on the neutron star EoS.

Instead of measuring the binary parameters $\boldsymbol{\theta}$ for individual events, we are interested in measuring the hyperparameters, $\boldsymbol{\Lambda} = (\boldsymbol{\Lambda}_{\mathrm{GW}}, \boldsymbol{\Lambda}_{\mathrm{EoS}})$ governing the distributions of $\boldsymbol{\theta}$ across a population of sources. The likelihood of observing $d, M_{\mathrm{rem}}$ given $\boldsymbol{\Lambda}$ is obtained by marginalizing over $\boldsymbol{\theta}$, and the likelihood of observing a set of events with $\{d, M_{\mathrm{rem}}\}$ is the product of the individual-event likelihoods:
\begin{align}
&p(d, M_{\mathrm{rem}} | \boldsymbol{\Lambda}_{\mathrm{GW}}, \boldsymbol{\Lambda}_{\mathrm{EoS}}) = \int p(d, M_{\mathrm{rem}} | \boldsymbol{\theta}, \boldsymbol{\Lambda}_{\mathrm{EoS}}) \pi_{\mathrm{pop}}(\boldsymbol{\theta} | \boldsymbol{\Lambda}_{\mathrm{GW}})d\boldsymbol{\theta}\\ \nonumber
&p(\{d, M_{\mathrm{rem}} \}| \boldsymbol{\Lambda}_{\mathrm{GW}}, \boldsymbol{\Lambda}_{\mathrm{EoS}}) \\
&\quad = \prod_{i}\int p(d_i | \boldsymbol{\theta}_i )\delta(\hat{M}_{\mathrm{rem}}(\mathbf{x}_i, \boldsymbol{\Lambda}_{\mathrm{EoS}})) \pi_{\mathrm{pop}}(\boldsymbol{\theta}_i | \boldsymbol{\Lambda}_{\mathrm{GW}})d\boldsymbol{\theta}.
\label{eq:joint_int}
\end{align}

The gravitational-wave likelihood in  Eq.~\ref{eq:joint_int} can be replaced via Bayes' Theorem with the ratio of the posterior to the prior, which allows the integral to be evaluated using a sum over individual-event posterior samples, $j$:
\begin{align}
\nonumber
    &p(\{d, M_{\mathrm{rem}} \}| \boldsymbol{\Lambda}_{\mathrm{GW}}, \boldsymbol{\Lambda}_{\mathrm{EoS}}) \\
&\quad \propto \prod_{i}\int \frac{p(\boldsymbol{\theta}_i | d_i)}{\pi_{\mathrm{PE}}(\boldsymbol{\theta}_i)} \delta(\hat{M}_{\mathrm{rem}}(\mathbf{x}_i, \boldsymbol{\Lambda}_{\mathrm{EoS}})) \pi_{\mathrm{pop}}(\boldsymbol{\theta}_i | \boldsymbol{\Lambda}_{\mathrm{GW}})d\boldsymbol{\theta}\\
&\quad \propto \prod_{i}\sum_{j} \frac{\delta(\hat{M}_{\mathrm{rem}}(\mathbf{x}_{i,j}, \boldsymbol{\Lambda}_{\mathrm{EoS}})) \pi_{\mathrm{pop}}(\boldsymbol{\theta}_{i,j} | \boldsymbol{\Lambda}_{\mathrm{GW}})}{\pi_{\mathrm{PE}}(\boldsymbol{\theta}_{i,j})}
\label{eq:join_like_noVT}
\end{align}

Because we are neglecting electromagnetic selection effects, the joint likelihood in Eq.~\ref{eq:join_like_noVT} can be amended to account for gravitational-wave selection effects in the usual way,
\begin{align}
\nonumber
    &p(\{d, M_{\mathrm{rem}} \}| \boldsymbol{\Lambda}_{\mathrm{GW}}, \boldsymbol{\Lambda}_{\mathrm{EoS}}) \\
&\quad \propto \frac{1}{\alpha(\boldsymbol{\Lambda}_{\mathrm{GW}})^{N}}\prod_{i}^{N}\sum_{j} \frac{\delta(\hat{M}_{\mathrm{rem}}(\mathbf{x}_{i,j}, \boldsymbol{\Lambda}_{\mathrm{EoS}})) \pi_{\mathrm{pop}}(\boldsymbol{\theta}_{i,j} | \boldsymbol{\Lambda}_{\mathrm{GW}})}{\pi_{\mathrm{PE}}(\boldsymbol{\theta}_{i,j})}
\label{eq:join_like}
\end{align}
where $N$ is the number of observed NSBH mergers in the population being analyzed. This is the same expression as Eq.~\ref{eq:hyper-like} with the addition of the remnant mass likelihood to the numerator.

\section{Black hole spin comparison}
\label{ap:bh_spin}
In our analysis, we use uniform priors on the $\alpha_{\chi}, \beta_\chi$ hyperparameters governing the black hole spin magnitude Beta distribution and include values of $\alpha_{\chi}, \beta_\chi \leq 1$, which correspond to singular Beta distributions. This choice allows the spin distribution to peak at $\chi_{\mathrm{BH}}=0$, as we might expect if NSBHs form via isolated binary evolution and the black hole is the first compact object to form~\citep{Qin:2018vaa, Bavera:2021evk, Chattopadhyay:2020lff, 2019MNRAS.485.3661F, Spruit:2001tz}, or $\chi_{\mathrm{BH}}=1$, consistent with measurements of spin in black hole X-ray binaries~\citep[e.g.,][]{Reynolds:2020jwt}. However, in order to obtain an apples-to-apples comparison between the black hole spin distribution we infer for NSBH and the one measured in \cite{LIGOScientific:2021psn}, we need to apply the same prior. The LVK analysis does not allow for singular Beta distributions, and instead applies uniform priors on the mean and variance of the Beta distribution,
\begin{align}
    \mu &= \frac{\alpha}{\alpha + \beta}\\
    \sigma^{2} &= \frac{\alpha\beta}{(\alpha+\beta)^{2}(\alpha + \beta + 1)}.
\end{align}

\begin{figure}
	\centering
	\includegraphics[width=0.85\columnwidth]{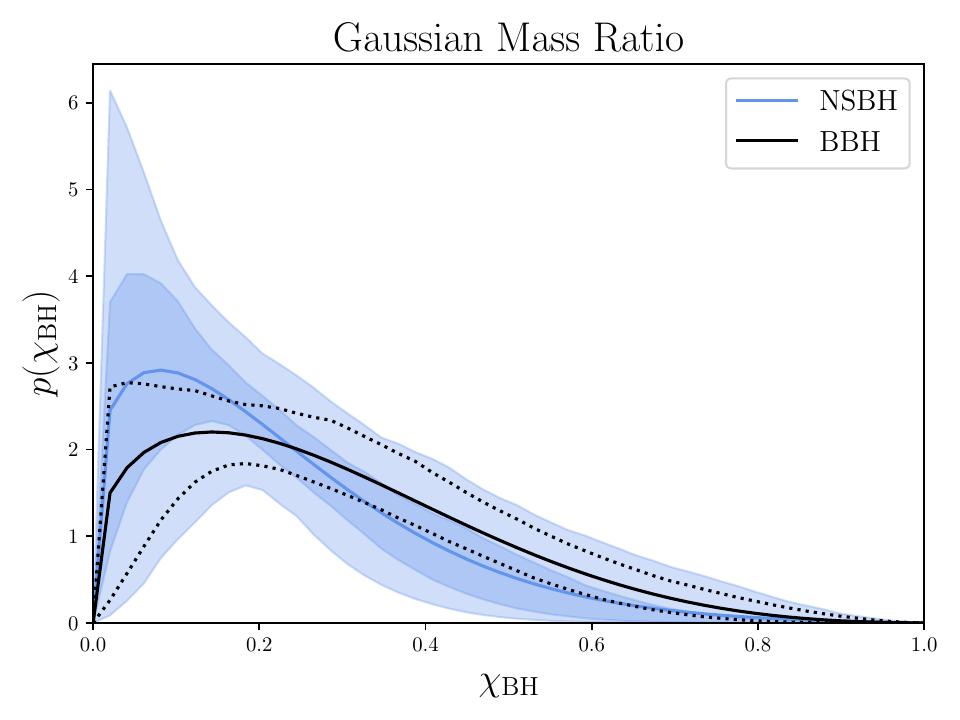}
	\includegraphics[width=0.85\columnwidth]{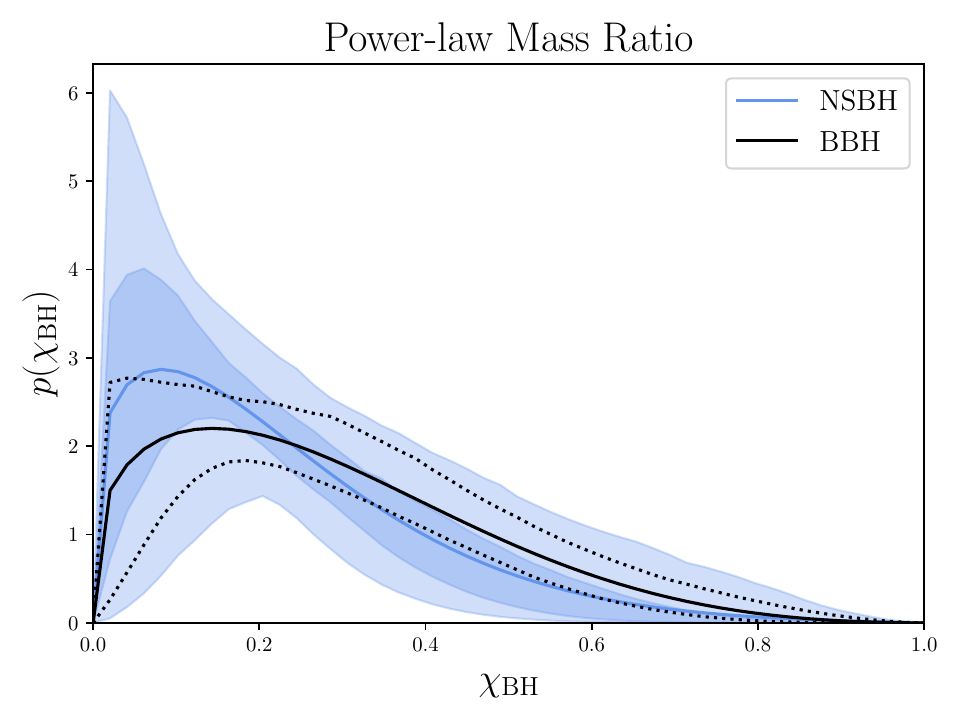}
    \caption{Posterior on the black hole spin magnitude distribution for the NSBH population obtained using gravitational-wave data alone (blue) using the same prior assumptions that go into the BBH spin magnitude distribution inference in \protect\cite{LIGOScientific:2021psn}, shown in black. The shaded blue region shows the NSBH 50\% and 90\% credible intervals, while the dotted black lines enclose the 90\% credible region for BBH. The top (bottom) shows the result under the Gaussian (power-law) pairing function.}
    \label{fig:spin_comp}
\end{figure}

In Fig.~\ref{fig:spin_comp}, we show in blue the posterior on the NSBH black hole spin magnitude distribution obtained using GW data alone (bottom left panels of Fig.~\ref{fig:ppds}) reweighted to match the prior choices made in the LVK analysis of BBH spins, while the BBH spin distribution is shown in black. This direct comparison demonstrates that our conclusion that the black holes in NSBHs have smaller spins still holds when equivalent prior assumptions are made. The NSBH result is considerably more uncertain than the BBH result, consistent with the fact that there are only four NSBH events included in our analysis but 69 BBH events going into the LVK result. With this choice of prior, we obtain $\chi_{\mathrm{BH},99} = 0.57^{+0.27}_{-0.17}$ ($\chi_{\mathrm{BH},99} = 0.53^{+0.34}_{-0.10}$) for the Gaussian (power-law) pairing function, while for the BBH analysis, $\chi_{\mathrm{BH},99} = 0.76^{+0.06}_{-0.06}$. %

\section{Analysis of the two most significant events}
Here we present the population distributions for the component masses, mass ratio, and black hole spin magnitude inferred using only the two lowest-FAR events, GW200105\_162426 and GW200115\_042309. By comparison with Fig.~\ref{fig:ppds}, the distributions obtained with just these two events are similar to those obtained including the less significant candidates, just more uncertain.
\begin{figure}
	\includegraphics[width=\columnwidth]{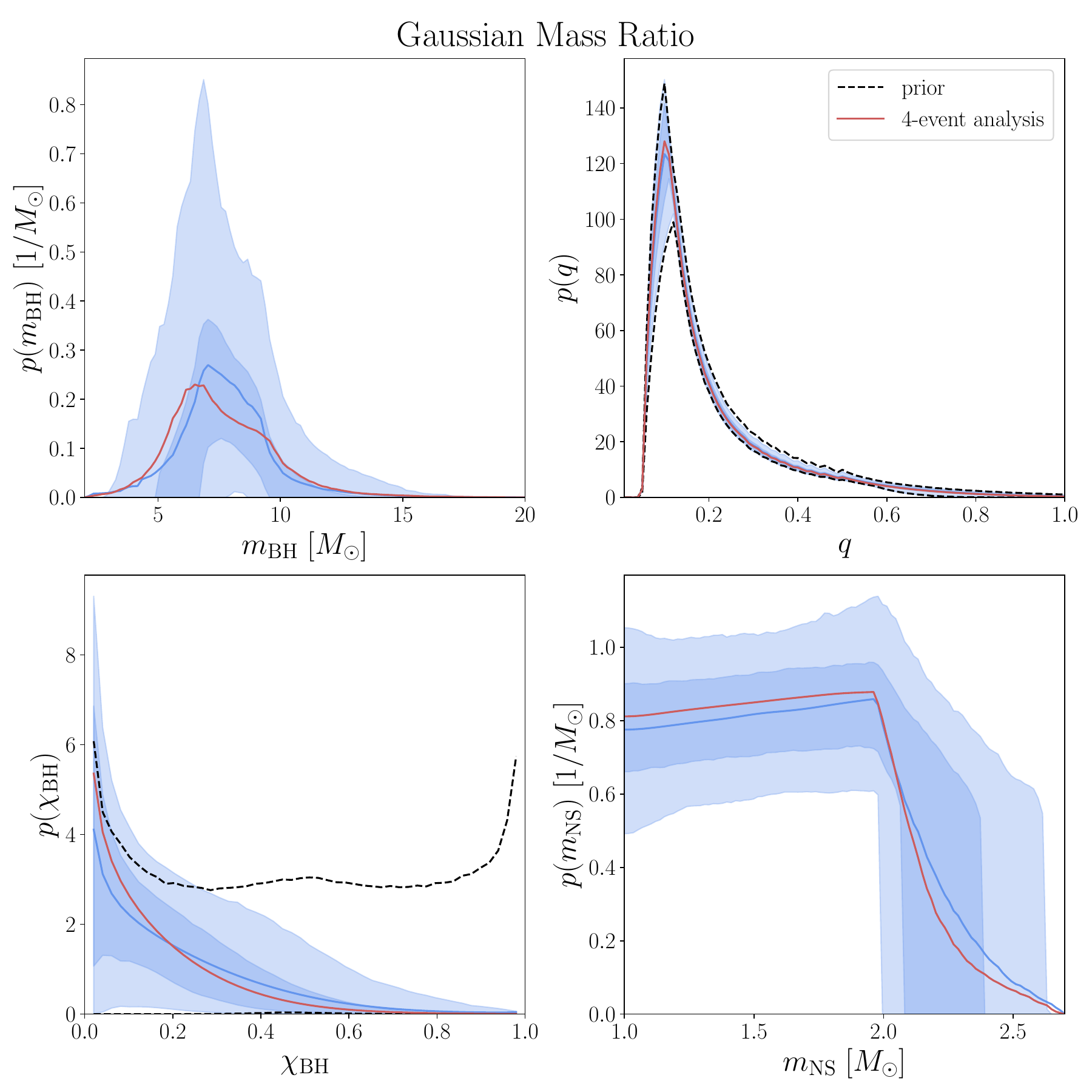}
    \caption{Posterior predictive distributions (solid blue) and 50\% and 90\% credible intervals (shaded blue) for the component masses, mass ratio, and black hole spin in the underlying, astrophysical NSBH population under the Gaussian mass ratio model using a population consisting of only GW200105\_162426 and GW200115\_042309. The black dashed lines show the 90\% credible region enclosed by draws from the hyperparameter prior for the black hole spin and mass ratio, and the red lines show the PPDs inferred using the original event selection for Fig.~\ref{fig:ppds}.}
    \label{fig:low_far}
\end{figure}

\section{Including GW190425}
\label{ap:gw190425}
In order to verify the effect of our event selection and imposed prior knowledge of source classification on our results, we repeat our analysis of only the GW data to include GW190425, widely considered to be a binary neutron star merger given its inferred masses. This system has posterior support in the region of parameter space covered by our hyperparameter priors in Table~\ref{tab:priors}, so it is not a priori excluded from our population, as is the case for GW190814.

Because this source is believed to be a binary neutron star rather than a NSBH merger, all of the publicly-available posterior samples for this event use waveform models that assume the tidal deformabilities of the two components are free parameters but do not include the effect of higher-order modes. This is in contrast to the posteriors we had analyzed for the analysis in the main text obtained with waveforms that assume both components are point masses with no tidal deformability but include the effects of higher-order modes. We choose to use the \texttt{PublicationSamples} posteriors for GW190425, where the spin magnitude prior extends up to $\chi < 0.89$ and the directions can be misaligned to the orbital angular momentum. We cannot self-consistently perform the analysis that models the NS EoS using the electromagnetic counterpart nondetection including GW190425, because there are no available posterior samples assuming the two components are point masses.

\begin{figure}
	\includegraphics[width=\columnwidth]{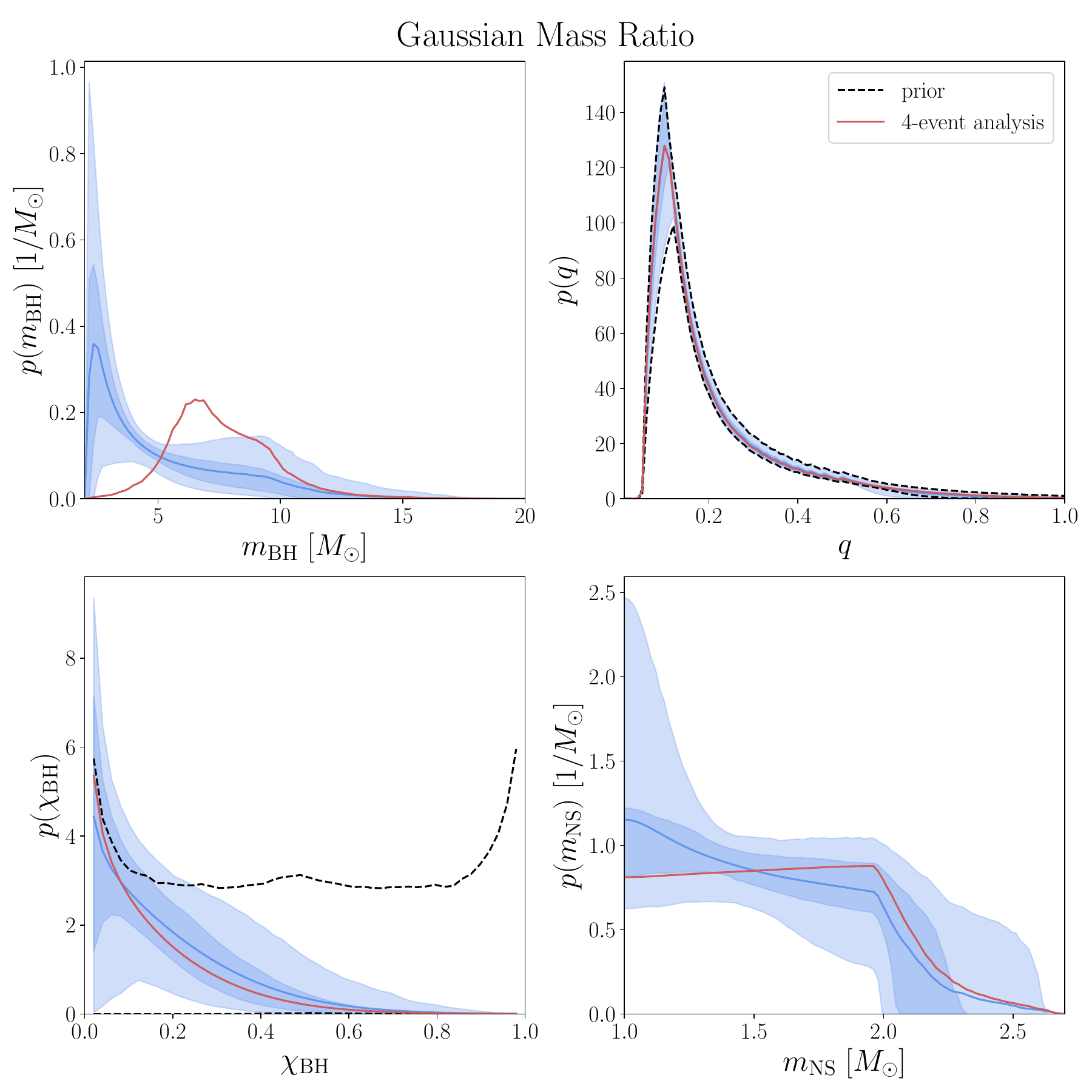}
    \caption{Posterior predictive distributions (solid blue) and 50\% and 90\% credible intervals (shaded blue) for the component masses, mass ratio, and black hole spin in the underlying, astrophysical NSBH population under the Gaussian mass ratio model using a population consisting of the four candidates analyzed in the main text \textit{and} GW190425. The black dashed lines show the 90\% credible region enclosed by draws from the hyperparameter prior for the black hole spin and mass ratio, and the red lines show the PPDs inferred using the original event selection for Fig.~\ref{fig:ppds}.}
    \label{fig:ppds_190425}
\end{figure}

\begin{figure}
	\includegraphics[width=\columnwidth]{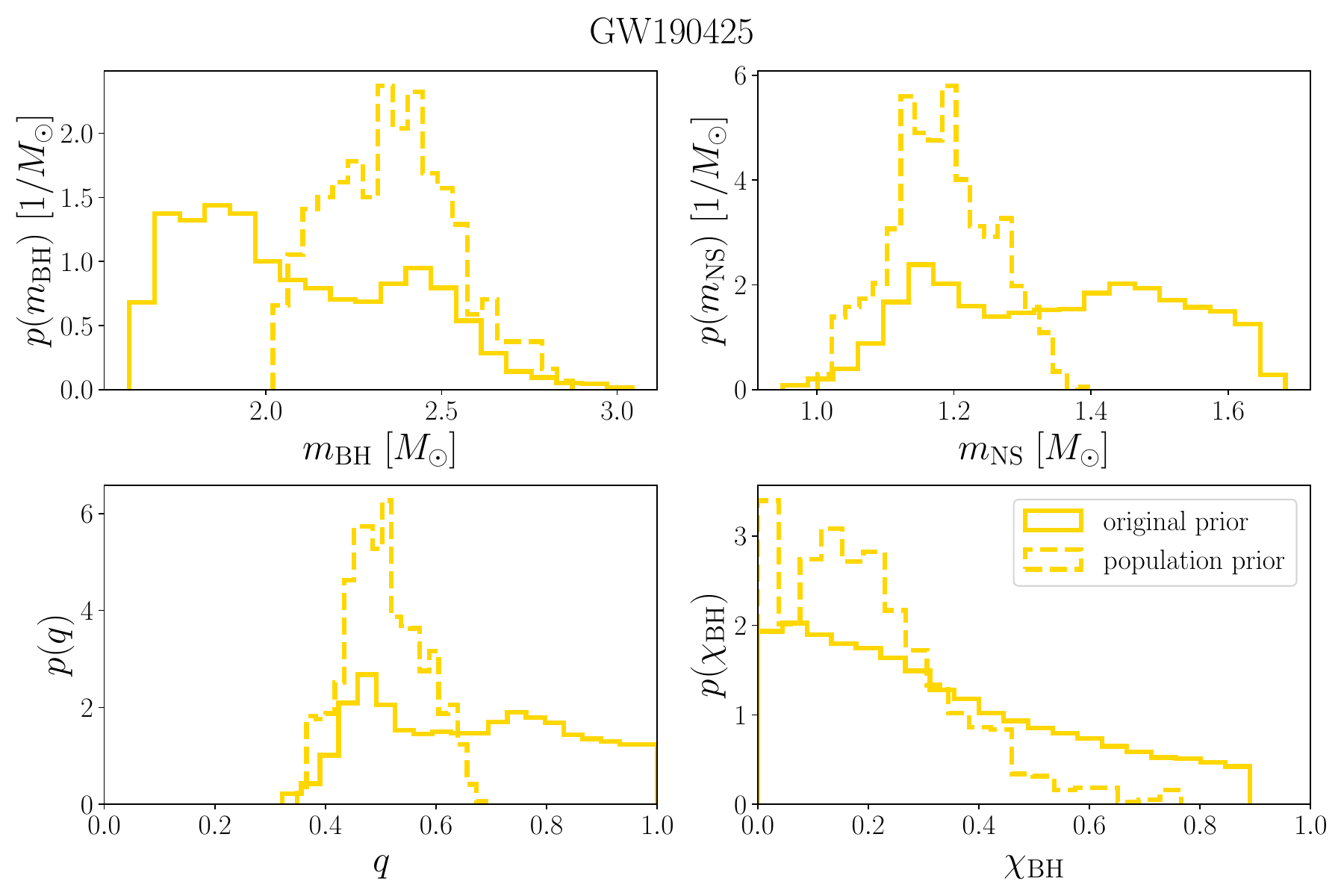}
    \caption{Distributions of the component masses, mass ratio, and black hole spin for GW190425 assuming it belongs to our population NSBH mergers under the original priors (solid lines) and reweighted into the population prior inferred under the Gaussian mass ratio model (dashed line).}
    \label{fig:1d_hists_190425}
\end{figure}

\begin{figure}
	\includegraphics[width=\columnwidth]{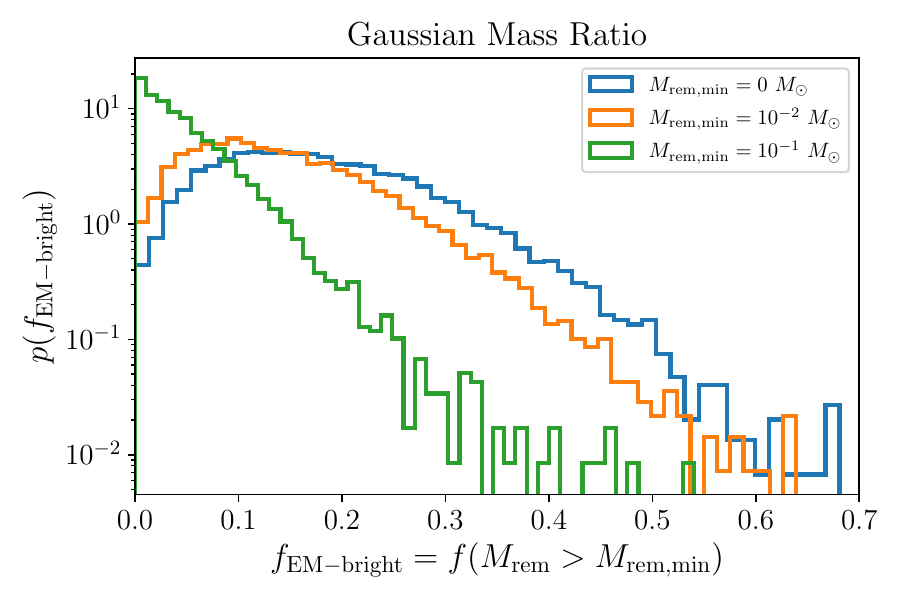}
    \caption{Posterior on the fraction of GW-detectable NSBH systems that will be electromagnetically bright with remnant mass $\hat{M}_{\mathrm{rem}} \geq M_{\mathrm{rem, min}}$ obtained by analyzing a population consisting of GW190425 along with the four candidate events presented in the main text. The different colors indicate different values of $M_{\mathrm{rem, min}}$. The posterior is marginalized over the uncertainty in the neutron star equation of state and in the population hyperparameters for the Gaussian pairing function.}
    \label{fig:em_bright_190425}
\end{figure}

The results of the analysis including GW190425 are presented in Figs.~\ref{fig:ppds_190425}-\ref{fig:em_bright_190425}. As might be expected given the low primary mass inferred individually for GW190425, the posterior predictive distribution for the black hole mass changes significantly when this event is included in the analysis, compared to the distribution presented in Fig.~\ref{fig:ppds}. The posterior on the minimum black hole mass rails against the lower edge of the prior, shifting the PPD to peak at $2~M_{\odot}$ rather than $\sim 6~M_{\odot}$. The upper tail of the distribution does not change substantially when this event is included, consistent with the fact that the posteriors on the maximum black hole mass are qualitatively similar with or without GW190425. The posterior on $\alpha$, however, more strongly disfavors negative values when GW190425 is included, as the peak at low black hole masses required by GW190425 must be accommodated by a power-law with a negative slope (positive $\alpha$).

The marginalized mass ratio distribution does not vary significantly upon including GW190425 in the analysis, as it is prior-driven, similar to the result obtained without GW190425. However, the posteriors on the mass ratio mean and width more strongly prefer values at the lower edges of their respective priors. This may be due to the information gained by increasing the number of events analyzed by 25\%. The distributions of the black hole spin and neutron star mass also do not change significantly when GW190425 is included.

The posteriors for the binary parameters of GW190425 under the original prior and reweighted into the new inferred population prior are shown in Fig.~\ref{fig:1d_hists_190425}. The effect of assuming that this event is an NSBH rather than a binary neutron star merger is to upweight the posterior support at unequal mass ratios. All the support at equal mass is removed, and the mass ratio posterior under the population prior peaks staunchly at $q=0.49$. The spin distribution also shifts towards lower values.

Finally, in Fig.~\ref{fig:em_bright_190425}, we show the posteriors on the fraction of GW-detectable NSBH systems that will be EM-bright for different values of the threshold remnant mass obtained including GW190425. As expected given the shift in the black hole mass distribution towards lower masses which leads to more significant disruption of the neutron star outside the ISCO radius, there is a commensurate increase in the EM-bright fraction relative to the result without GW190425 shown in Fig.~\ref{fig:em_bright}. When GW190425 is included, the posterior peaks away from $f_{\mathrm{EM-bright}}=0$, and we infer $f(\hat{M}_{\mathrm{rem}} > 0~M_{\odot}) = 0.16^{+0.13}_{-0.16}$.

While the posteriors for the minimum black hole mass and EM-bright fraction change significantly when GW190425 is included in the analysis, our main conclusions that the black holes in NSBHs are both less massive and more slowly spinning than those in BBH remain robust. The change in these posteriors highlights the importance of event selection and prior knowledge of source classification in population analyses. As suggested in the main text, this dependence can be removed by simultaneously fitting and sorting individual events into multiple sub-populations based on distinct distributions~\citep{Farah:2021qom}. We leave this analysis to future work.

\bsp	%
\label{lastpage}
\end{document}